\documentclass[sn-nature]{sn-jnl}% Style for submissions to Nature Portfolio journals
\usepackage{graphicx}%
\usepackage{multirow}%
\usepackage{amsmath,amssymb,amsfonts}%
\usepackage{amsthm}%
\usepackage{mathrsfs}%
\usepackage[title]{appendix}%
\usepackage{xcolor}%
\usepackage{textcomp}%
\usepackage{manyfoot}%
\usepackage{booktabs}%
\usepackage{algorithm}%
\usepackage{algorithmicx}%
\usepackage{algpseudocode}%
\usepackage{listings}%
\usepackage{bm}
\usepackage{siunitx}
\usepackage{framed}
\usepackage{braket}
\usepackage{soul}

\usepackage[T1]{fontenc}
\usepackage[latin9]{inputenc}
\usepackage{longtable}
\usepackage{amsmath}
\usepackage{amssymb}
\usepackage{graphicx}
\usepackage{esint}
\usepackage{babel}
\usepackage{afterpage,lscape}

\theoremstyle{thmstyleone}%
\theoremstyle{thmstyletwo}%

\theoremstyle{thmstylethree}%

\begin{document}

\title[Article Title]{$\frac{1}{2}\mathbb{Z}$ Topological Invariants and the Half Quantized
Hall Effect}


\author*[1]{\fnm{Fu} \sur{Bo}}\email{fubo@gbu.edu.cn}

\author*[2,3]{\fnm{Shen} \sur{Shun-Qing}}\email{sshen@hku.hk}

\affil[1]{\orgdiv{School of Sciences}, \orgname{Great Bay University}, \orgaddress{\city{Dongguan}, \postcode{523000}, \state{Guangdong Province}, \country{China}}}

\affil[2]{\orgdiv{Department of Physics}, \orgname{The University of Hong Kong}, \orgaddress{\street{Pokfulam Road}, \city{Hong Kong},   \country{China}}}

\affil[3]{\orgname{Quantum Science Center of Guangdong-Hong Kong-Macau Greater Bay Area},   \country{China}}


\abstract{The half-quantized Hall phase represents a unique metallic or semi-metallic
state of matter characterized by a fractional quantum Hall conductance,
precisely half of an integer $\nu$ multiple of $e^{2}/h$. Here we
demonstrate the existence of a novel $\frac{1}{2}\mathbb{Z}$ topological
invariant that sets the half-quantized Hall phase apart from two-dimensional
ordinary metallic ferromagnets. The $\frac{1}{2}\mathbb{Z}$ classification
is determined by the line integral of the intrinsic anomalous Hall
conductance, which is safeguarded by two distinct categories of local
unitary and anti-unitary symmetries in proximity to the Fermi surface
of electron states. We further validate the $\frac{1}{2}\mathbb{Z}$
topological order in the context of the quantized Hall phase by examining
semi-magnetic topological insulator $\mathrm{Bi}_{2}\mathrm{Te}_{3}$
and $\mathrm{Bi}_{2}\mathrm{Se}_{3}$ film for $\nu=1$ and topological
crystalline insulator SnTe films for $\nu=2$ or $4$. Our findings
pave the way for future exploration and understanding of topological
metals and their unique properties.}




\maketitle

\section{Introduction}\label{sec1}

The classification of the electronic states according to topological
invariants has revolutionized the understanding of quantum matter
and the underlying quantum behaviors \cite{schnyder2008classification,kitaev2009periodic,ryu2010topological,morimoto2013topological,zhao2013topological,matsuura2013protected,shiozaki2014topology,chiu2014classification,chiu2016classification}.
The $\mathbb{Z}$ and $\mathbb{Z}_{2}$ classifications have provided
a robust framework for distinguish the topological insulating phases
and superconducting phases from the many-body phases with the bulk
energy gap \cite{hasan2010colloquium,qi2011topological,shen2017topological}.
The bulk-boundary correspondence, which lies at the heart of the topological
insulating phases, asserts that the topological invariants of a material's
bulk dictate the characteristics of its boundary states, leading to
the discovery of a series of quantum phenomena and quantum materials
in condensed matter.

Integer quantum Hall effect was the first to be characterized by the
$\mathbb{Z}$ topological invariant in two dimensions. In their pioneering
work, Thouless, Kohmoto, Nightingale and den Nijs (TKNN) \cite{thouless1982quantized,niu1985quantize}
identified the integer $\nu$ in quantum Hall conductivity $\sigma_{H}=\nu\frac{e^{2}}{h}$
as an topological invariant. The integer is given by an integral of
the Bloch wave function over the Brillouin zone (BZ). Because fo its
periodic boundary condition, the first Brillouin zone of the electronic
band is equivalent to the surface of a torus, the integer is identified
to be the first Chern number of the surface curvature of the torus
\cite{Simon-83prl}. This number sets an integer quantum Hall phase
and quantum anomalous Hall phase from a simple band insulator. The
Chern number is closely associated with the number of the gapless
edge states on the system boundary, leading to the bulk-edge correspondence
for the topological phases \cite{hatsugai1993chern}. Theoretical
prediction and experimental observation of quantum anomalous Hall
effect were one of the successful examples to explore the quantum
state of matter in condensed matter \cite{haldane1988model,yu2010quantized,chu2011surface,chang2013experimental,checkelsky2014trajectory}.

Further more, the nonzero Hall conductivity breaks the time reversal
symmetry. The Chern number must vanishes in a time-reversal invariant
system. A $\mathbb{Z}_{2}$ topological invariant was introduced by
Kane and Mele \cite{kane2005z} to characterize quantum spin Hall
effect in the systems with spin-orbit coupling. A $\mathbb{Z}_{2}$
index means the appearance of a pair of helical edge states in the
band gap around the system boundary. This is robust against the perturbation
against interaction and disorder \cite{bernevig2006hgte}. Thus the
$\mathbb{Z}$ and $\mathbb{Z}_{2}$ topological invariant distinguish
the topological phases from an ordinary insulator.

The anomalous Hall effect in ferromagnets usually gives rise to a
non-quantized Hall conductivity, which can be expressed by an integral
of the Berry curvature in the Brillouin zone. Haldane \cite{haldane2004berry}
found that the anomalous Hall conductivity in a metallic ferromagnet
is expressible in terms of the Berry phase of the quasiparticles moving
on the Fermi surface. In an insulating ferromagnet, the band gap provides
a protection on the robustness of quantization of the anomalous Hall
conductivity. Due to nonexistence of an band gap, this raises a question
whether the anomalous Hall conductivity can be quantized in a metallic
ferromagnet. Fu et al \cite{fu2022quantum} proposed quantum anomalous
semimetal, which consists of gapless Wilson fermions and is characterized
a one half topological invariant, or the half quantized Hall conductivity.
Mogi et al \cite{mogi2022experimental} measured one half quantum
Hall conductivity in a semi-magnetic topological insulator film, and
suggested a signature of the parity anomaly of massive Dirac fermions.
Zou et al \cite{zou2022half,zou2023half} analyzed the band structure
of the magnetic topological insulator and found the existence of the
gapless Dirac fermions, which is attributed to the quantization of
the Hall conductivity. Since then there are many efforts to understand
the effect \cite{gong2023half,yang2023realization,ning2023robustness,wang2024signature,wan2024quarter,zhou2024dissipative}.

In this article, we introduce the $\frac{1}{2}\mathbb{Z}$ topological
invariant to distinguish the half quantized Hall phase from an ordinary
metallic ferromagnet, which is enforced by the local symmetry near
the Fermi surface. The anomalous Hall conductivity can be expressed
in terms of the Berry connection, 
\begin{equation}
\sigma_{xy}=\frac{e^{2}}{2\pi h}\sum_{i}\int_{\mathrm{C_{FS}^{i}}}d\mathbf{l}\cdot\mathbf{A}_{i}(\mathbf{k})\label{eq:loop_formula}
\end{equation}
where the summation over $i$encompasses all the separate curves of
the Fermi surfaces, and $\mathbf{A}_{i}(\mathbf{k})$ represents the
corresponding Berry connection for each curve $i$. The formula in
Eq. (\ref{eq:loop_formula}) is strictly on the states at the Fermi
surface ($\mathbf{k}_{F}\in C_{FS}^{i}$). In a partially filled two-dimensional
band, if the electron states near the Fermi surface are invariant
under a specific symmetry, the intrinsic Hall conductance has to be
quantized to be half of an integer $\nu$ multiple of $e^{2}/h$,
i.e., $\sigma_{H}=\frac{\nu}{2}\frac{e^{2}}{h}$. We identified two
distinct categories of local unitary symmetries ($\sigma_{v}$ and
$C_{2x}$) and five types of local anti-unitary symmetries ( $T$,$C_{nz}T$,
$IT$, $\sigma_{h}T$ and $S_{nz}T$ ) as listed in Table 1, leading
to a $\frac{1}{2}\mathbb{Z}$ topological invariant to characterize
the topological metallic or semi-metallic state of matter. For a nonzero
$\nu$, the global symmetry must be broken for the whole system. The
$\frac{1}{2}\mathbb{Z}$ classification, which is based solely on
the symmetry of states at the Fermi surfaces, is distinct from the
$\mathbb{Z}$ and $\mathbb{Z}_{2}$classifications for insulators
and superconductors that depend on global symmetries.

This leads us to introduce the concept of 'local symmetry', where
a symmetry operation is applied only to a subspace of the entire Brillouin
zone. The local symmetry may occur accidentally such that the Hall
conductivity equals to one half of $\frac{e^{2}}{h}$. In magnetic
or semi-magnetic topological (crystalline) insulators, the local symmetry
of the surface states facilitates the emergence of the Hall conductivity
plateaus as the chemical potential shifts within local-symmetry-preserving
regions of the Brillouin zone \cite{zou2023half}. This observation
implies that the quantized Hall metallic phase is stable and not just
a transient point between two insulating phases. Our theoretical framework
and computational evidence further substantiate the existence of half-quantized
Hall conductivity in semi-magnetic heterostructure of topological
insulator $\mathrm{Bi}_{2}\mathrm{Te}_{3}$ and $\mathrm{Bi}_{2}\mathrm{Se}_{3}$
and topological crystalline insulator SnTe films. We provide a comprehensive
classification of topological surface states, highlighting how local
symmetries affect the electronic properties of gapless Dirac cones.
Disruption of these symmetries in semi-magnetic films leads to variable
half-quantized Hall conductivities. Validation through tight-binding
model calculations supports our findings and suggests new experimental
pathways. Moreover, we explore how extrinsic mechanisms are marginalized
by the presence of local symmetries, thus stabilizing the Hall effect
in novel semi-magnetic metals.

This paper is organized as follows:  We begin with a preliminary introduction to the Fermi-loop formula for Hall conductivity. Next, we present a proof of the half-quantization of Hall conductivity and identify the corresponding local symmetries. Subsequently, we delve into the distinctions between local and global symmetry breaking. Following this, we classify a single gapless Dirac cone with symmetry $C_{nz}T$ on a lattice. We then discuss two materials of topological insulators,  $\mathrm{Bi}_{2}\mathrm{Te}_{3}$ and $\mathrm{Bi}_{2}\mathrm{Se}_{3}$ 
for $\nu=1$ and the topological crystalline insulator SnTe for $\nu=2$ or
$4$. Finally, we explore the disappearance of the extrinsic anomalous Hall effect in semi-magnetic topological insulators

\section{Results}\label{sec2}

\paragraph*{Fermi-loop formula for Hall conductivity}

We initiate our discussion with the Kubo-Streda formula \cite{kubo1957statistical,streda1982theory},
which posits that the electric conductivity tensor $\sigma_{\alpha\beta}$
can be decomposed into $\sigma_{\alpha\beta}=\sigma_{\alpha\beta}^{Ia}+\sigma_{\alpha\beta}^{Ib}+\sigma_{\alpha\beta}^{II}$
with
\begin{align*}
\sigma_{\alpha\beta}^{Ia} & =-\frac{e^{2}\hbar}{2\pi V}\int d\varepsilon\partial_{\varepsilon}f(\varepsilon)\mathrm{Tr}[\hat{v}_{\alpha}G^{R}\hat{v}_{\beta}G^{A}],\\
\sigma_{\alpha\beta}^{Ib} & =\frac{e^{2}\hbar}{4\pi V}\int d\varepsilon\partial_{\varepsilon}f(\varepsilon)\mathrm{Tr}[\hat{v}_{\alpha}G^{R}\hat{v}_{\beta}G^{R}+\hat{v}_{\alpha}G^{A}\hat{v}_{\beta}G^{A}],\\
\sigma_{\alpha\beta}^{II} & =\frac{e^{2}\hbar}{2\pi V}\int d\varepsilon f(\varepsilon)\mathrm{Re}\mathrm{Tr}[\hat{v}_{\alpha}G^{R}\hat{v}_{\beta}\partial_{\varepsilon}G^{R}-\hat{v}_{\alpha}\partial_{\varepsilon}G^{R}\hat{v}_{\beta}G^{R}].
\end{align*}
In the expressions, $\hat{v}_{\alpha}$ are the velocity operators
along $\alpha$ direction, $G^{R(A)}$ represents the retarded (advanced)
Green's function of the system, and the energy argument has been omitted
for the sake of brevity. The function $f(\varepsilon)$ denotes the
Fermi-Dirac distribution, and $\partial_{\varepsilon}$ signifies
a derivative with respect to energy $\varepsilon$. Owing to their
link to $\partial_{\varepsilon}f(\varepsilon)$ and $f(\varepsilon)$,
$\sigma_{\alpha\beta}^{I}$ and $\sigma_{\alpha\beta}^{II}=e\frac{\partial n(\varepsilon)}{\partial B}|_{\epsilon=\mu,B=0}$
are frequently referred to as the Fermi surface and Fermi sea terms,
respectively. In the clean limit, where disorder effects are minimal,
the Hall response is primarily due to the intrinsic mechanisms. Specifically,
it stems only from $\sigma_{\alpha\beta}^{II}$ and $\sigma_{\alpha\beta}^{I}$
vanishes when the chemical potential $\mu$ lies within the energy
gap, whereas it arises from both $\sigma_{\alpha\beta}^{I}$ and $\sigma_{\alpha\beta}^{II}$
when the chemical potential crosses into the band. When impurity scattering
becomes significant, extrinsic contributions such as side jump and
skew scattering which comes from $\sigma_{\alpha\beta}^{Ia}$ after
disorder averaging\cite{sinitsyn2007anomalous,nagaosa2010anomalous},
cannot be ignored. Here, we focus on the intrinsic contributions.

For a two-dimensional electronic system, taking into account the expansion
of the trace in the basis of Bloch states $\{|u_{n\mathbf{k}}\}$,
the Hall conductivity can be represented as\cite{xiao2009berryphase}
\[
\sigma_{xy}=\frac{e^{2}}{\hbar}\sum_{n}\int_{\mathrm{Bz}}\frac{d\mathbf{k}}{(2\pi)^{2}}f_{n}(\mathbf{k})\Omega_{xy}^{n}(\mathbf{k}),
\]
where the integral is carried out over the two-dimensional Brillouin
zone. The function $f_{n}(\mathbf{k})$ denotes the occupation function
and the summation runs over those states that are occupied. $\Omega_{xy}^{n}(\mathbf{k})$
is the Berry-curvature of band $n$, defined as $\Omega_{xy}^{n}=-2\mathrm{Im}\langle\frac{\partial u_{n\mathbf{k}}}{\partial k_{x}}|\frac{\partial u_{n\mathbf{k}}}{\partial k_{y}}\rangle$.
This result connects the Fermi sea transport contributions to the
Berry curvature of the Bloch states which has been widely used to
characterize the intrinsic Hall effect of bulk materials.

Applying the Stokes' theorem, the areal integral of the Berry curvature
for the $n$th band over a specific region $\mathrm{S}_{n}$ in the
Brillouin zone can be converted into a loop integral of the Berry
connection $\mathbf{A}^{n}(\mathbf{k})$ around $\mathrm{C}_{n}$
, the curve that defines the oriented curve bounding $\mathrm{S}_{n}$:

\begin{equation}
\int_{\mathrm{S}_{n}}d^{2}\mathbf{k}\Omega_{xy}^{n}(\mathbf{k})=\int_{\mathrm{C}_{n}}d\mathbf{l}\cdot\mathbf{A}^{n}(\mathbf{k}).\label{eq:stokestheorem}
\end{equation}
For a fully filled band, if the Berry connection $\mathbf{A}^{n}(\mathbf{k})$
is smoothly defined throughout the entire Brillouin zone, then $\mathrm{S}_{n}$
can encompass the whole Brillouin zone. In two dimensions, the Brillouin
zone is topologically equivalent to a torus, which lacks a boundary.
Consequently, the Chern number for $n$th band $\nu^{n}=\int_{\mathrm{Bz}}d^{2}\mathbf{k}\Omega_{xy}^{n}(\mathbf{k})$
must be zero. Therefore, the presence of a non-zero Chern number presents
a \textquotedblleft topological obstruction\textquotedblright{} of
a globally smooth gauge \cite{thouless1984wannier,thonhauser2006insulator}.
If $\mathbf{A}^{n}(\mathbf{k})$ exhibits a singularity, the Brillouin
zone can be divided into two subregions, $\mathrm{S}_{n}^{1}$ and
$\mathrm{S}_{n}^{2}$, such that $\int_{\mathrm{Bz}}=\int_{\mathrm{S}_{n}^{1}}+\int_{\mathrm{S}_{n}^{2}}$.
Within each subregion, a smooth and continuous gauge is feasible.
Across the boundary of these subregions, the Berry connection differs
only by a gauge transformation, $\mathbf{A}_{1}^{n}(\mathbf{k})=\mathbf{A}_{2}^{n}(\mathbf{k})-\partial_{\mathbf{k}}\phi_{n}(\mathbf{k})$.
The oriented loop $\mathrm{C}_{n}$ forming the boundary traverses
$\mathrm{S}_{n}^{1}$ in the forward direction and $\mathrm{S}_{n}^{2}$
in the reverse direction. By utilizing Stokes' theorem in Eq. (\ref{eq:stokestheorem})
to each subregion and adding the results, we find $\nu^{n}=\int_{\mathrm{Bz}}d^{2}\mathbf{k}\Omega_{xy}^{n}(\mathbf{k})=\int_{\mathrm{C}_{n}}\partial_{\mathbf{k}}\phi_{n}(\mathbf{k})$.
Because $\phi_{n}(\mathbf{k})$ is a unique function on $\mathrm{C}_{n}$,
the resulting $\nu^{n}$ must be an integer. For a completely filled
band, where the Brillouin zone is regarded as a closed manifold without
boundary, the Chern number $\nu^{n}$ must be an integer, reflecting
the topological quantization inherent to the fully filled band structure
\cite{thouless1982quantized}. This leads to the $\mathbb{Z}$ classification
of the topological insulators in two dimensions.

For a partially filled band, the integral over the occupied states
forms an open manifold $S_{n}$ within the Brillouin zone, bounded
by the curve $\mathrm{C}_{n}$ corresponding to the sum of all the
loops of the Fermi surfaces. Invoking Stokes' theorem in Eq. (\ref{eq:stokestheorem}),
the Hall conductivity for this band can be recast as a loop integral
over the Fermi surfaces:
\begin{equation}
\sigma_{xy}^{n}=\frac{e^{2}}{2\pi h}\int_{\mathrm{C_{FS}}}d\mathbf{l}\cdot\mathbf{A}_{n}(\mathbf{k}).\label{eq:Berry_connection}
\end{equation}
In this situation $\int_{\mathrm{C_{FS}}}d\mathbf{l}\cdot\mathbf{A}_{n}(\mathbf{k})=\phi_{n}^{\mathrm{B}}$,
which has the interpretation of a Berry phase. This equation elucidates
the connection between the global property of the band structure encapsulated
by the Berry curvature and the local geometric phases described by
the Berry connection. It is critical to note that while determining
$\phi_{n}^{\mathrm{B}}$ using only the knowledge of the eigenstates
$|u_{n\mathbf{k}}\rangle$ on the curve $\mathrm{C}_{n}$, there will
exist a $2\pi$ ambiguity in comparison to the surface integral of
the Berry curvature. The equation only holds precisely when a gauge
choice is made that is smooth and continuous throughout the region
$S_{n}$, including its boundary $\mathrm{C}_{n}$, and this gauge
is employed to calculate the loop Berry phase. In Beenakker's tangent
fermion model\cite{beenakker2023tangent}, a singularity at $k=\pi$
precludes the straightforward application of the Berry connection
formula. This singularity introduces complications that require a
more nuanced approach to accurately describe the system's topological
properties.

Now we turn to the case of the multi-bands. We have assumed that each
band is isolated, meaning it remains separated by a finite energy
gap from both the adjacent lower and higher bands throughout the entire
Brillouin zone. However, in real crystals, the valence bands may become
degenerate at certain points or lines within the Brillouin zone due
to either symmetry or accidental degeneracy. Consequently, the Bloch
functions for energy eigenstates, often exhibit singularities as a
function of $\mathbf{k}$ near these degeneracies. To address this
situation, it is generally useful to define an isolated group of $J$
bands to be a set of $J$ consecutive energy bands that do not become
degenerate with any lower or higher band anywhere in the Brillouin
zone. The multi-band Berry connection and curvature for such a system
are given by: $\mathbf{A}^{mn}(\mathbf{k})=i\langle u_{m\mathbf{k}}|\boldsymbol{\partial}_{\mathbf{k}}|u_{n\mathbf{k}}\rangle$
and $\Omega_{xy}^{mn}(\mathbf{k})=\partial_{k_{x}}A_{y}^{mn}(\mathbf{k})-\partial_{k_{y}}A_{x}^{mn}(\mathbf{k})+i[A_{x}(\mathbf{k}),A_{y}(\mathbf{k})]^{mn}$.
The trace of these two quantities satisfies the same version of Stokes'
theorem as the single-band case in Eq. (\ref{eq:stokestheorem}).
The Hall conductivity resulting from the $J$ bands can be calculated
as
\[
\sigma_{xy}^{J}=\frac{e^{2}}{\hbar}\sum_{n=1}^{J}\int_{\mathrm{Bz}}\frac{d\mathbf{k}}{(2\pi)^{2}}f_{n}(\mathbf{k})\Omega_{xy}^{nn}(\mathbf{k}).
\]
In particular, if we assume the Fermi energy intersects a single band
$n=1$, and the other $J-1$ bands are fully occupied, we can divide
the Brillouin zone into two subregions $\mathrm{S}^{1}$ and $\mathrm{S}^{2}$:
$\mathrm{S}^{1}$ is the inner region enclosing by the Fermi surface
loop, where metallic band is well separated from all others ; $\mathrm{S}^{2}=\mathrm{BZ}-\mathrm{S}^{1}$,
the bands may experience degeneracies. Since the contribution comes
solely from the occupied states, the Hall conductivity can be reformulated
as an areal integral over these two subregions:
\begin{align*}
\sigma_{xy}^{J} & =\frac{e^{2}}{2\pi h}\left(\int_{\mathrm{S}^{1}}d\mathbf{k}\Omega_{xy}^{\mathrm{tr}^{\prime}}(\mathbf{k})+\int_{\mathrm{S}^{2}}d\mathbf{k}\Omega_{xy}^{\mathrm{tr}}(\mathbf{k})\right)
\end{align*}
where $\Omega_{xy}^{\mathrm{tr}}(\mathbf{k})=\sum_{n=1}^{J}\Omega_{xy}^{nn}(\mathbf{k})$
represents the trace over all $J$ bands, and $\Omega_{xy}^{\mathrm{tr}^{\prime}}(\mathbf{k})=\sum_{n=2}^{J}\Omega_{xy}^{nn}(\mathbf{k})$
excludes the band intersecting the Fermi surface. By applying the
Stokes' theorem, the Hall conductivity can be expressed as: 
\begin{align*}
\sigma_{xy}^{J} & =\frac{e^{2}}{2\pi h}\left(\int_{\mathrm{C}^{1}}d\mathbf{l}\cdot\mathbf{A}^{\mathrm{tr}}(\mathbf{k})+\int_{\mathrm{C}^{2}}d\mathbf{l}\cdot\mathbf{A}^{\mathrm{tr}^{\prime}}(\mathbf{k})\right)
\end{align*}
where $\mathbf{A}^{\mathrm{tr}}(\mathbf{k})=\sum_{n=1}^{J}\mathbf{A}^{nn}(\mathbf{k})$
and $\mathbf{A}^{\mathrm{tr}^{\prime}}=\sum_{n=2}^{J}\mathbf{A}^{nn}(\mathbf{k})$.
Along $\mathrm{C}^{1}$, where the metallic band is non-degenerate
with the others, we can express the trace of the Berry connection
as $\mathbf{A}^{\mathrm{tr}}(\mathbf{k})=\mathbf{A}^{\mathrm{tr}^{\prime}}(\mathbf{k})+\mathbf{A}^{11}(\mathbf{k})$.
Considering $\mathrm{C}^{1}$ and $\mathrm{C}^{2}$ are oriented in
opposite directions, the Hall conductivity is given by
\begin{equation}
\sigma_{xy}^{J}=\frac{e^{2}}{2\pi h}\int_{\mathrm{C}_{\mathrm{FS}}}d\mathbf{l}\cdot\mathbf{A}^{11}(\mathbf{k}).\label{eq:multiband_berry_connection}
\end{equation}
This equation summarizes the contribution to the Hall conductivity
from an isolated group of $J$ bands, considering the particular case
where the Fermi energy crosses only one band, and the remaining $J-1$
bands are fully occupied.

\paragraph*{Symmetry Constraint on Berry Curvature and Berry Connection}

The Hall conductivity is usually not quantized if the electronic bands
are partially filled or there exists a finite Fermi surface. Here
we show that if the states of electrons near the Fermi surface are
invariant under one of the specific symmetries listed in Table 1,
the intrinsic Hall conductance is quantized to be half of an integer
$\nu$(including zero) multiple of $e^{2}/h$, i.e.,
\begin{equation}
\sigma_{H}=\frac{\nu}{2}\frac{e^{2}}{h}.\label{eq:quantum-Hall}
\end{equation}
According to Wigner's theorem, all symmetry operators can be categorized into one of two types: unitary or anti-unitary symmetry \cite{wigner2012group}. We identified two distinct categories of local unitary symmetries
($\sigma_{v}$ and $C_{2x}$) and five types of local anti-unitary
symmetries ($T$, $C_{nz}T$,  $IT$,  $\sigma_{h}T$ and $S_{nz}T$), leading to a $\frac{1}{2}\mathbb{Z}$ topological invariant to
characterize the topological metallic or semi-metallic state of matter.
This classification parallels the $\mathbb{Z}$ and $\mathbb{Z}_{2}$
classifications for the insulating phases, and distinguishes the half-quantized Hall phase from the conventional two-dimensional metals or semimetals.  In the following, we prove the half quantization of the Hall conductivity under the protection of the unitary and anti-unitary symmetry, respectively.

\subparagraph*{The anti-unitary symmetry}

Consider a Bloch Hamiltonian $h(\mathbf{k})$ which is invariant under
an anti-unitary symmetry,\textit{ i.e.,} $S_{a}h(\mathbf{k})S_{a}^{-1}=h(D_{a}\mathbf{k}),$
where $D_{a}\mathbf{k}$ is the transformed wave vector $\mathbf{k}$
under $S_{a}.$ Due to this symmetry, the eigenstates of $h(\mathbf{k})$
at $\mathbf{k}$ and $D_{a}\mathbf{k}$ must be related by a gauge
transformation. Explicitly, for any eigenstate $|u_{n\mathbf{k}}\rangle$
of $h(\mathbf{k})$ with eigenvalue $\varepsilon_{n\mathbf{k}}$ we
have $h(D_{a}\mathbf{k})S_{a}|u_{n\mathbf{k}}\rangle=S_{a}h(\mathbf{k})|u_{n\mathbf{k}}\rangle=\varepsilon_{n\mathbf{k}}S_{a}|u_{n\mathbf{k}}\rangle.$
Thus $S_{a}|u_{n\mathbf{k}}\rangle$ is an eigenstate of $h(D_{a}\mathbf{k})$
with the same energy $\varepsilon_{n,D\mathbf{k}}=\varepsilon_{n\mathbf{k}}$.
We can thus expand $S_{a}|u_{n\mathbf{k}}\rangle=\sum_{m}B_{n,m}(\mathbf{k})|u_{mD_{a}\mathbf{k}}\rangle$
where $B_{n,m}(\mathbf{k})=\langle u_{mD_{a}\mathbf{k}}|S_{a}|u_{n\mathbf{k}}\rangle$
are the sewing matrix of the anti-unitary transformation which are
non-zero only when $\varepsilon_{n\mathbf{k}}=\varepsilon_{m,D_{a}\mathbf{k}}$.
We can factor out the complex conjugation $\mathcal{K}$ from $S_{a}$
as $S_{a}=U_{a}\mathcal{K}$, thereby defining the unitary operator
$U_{a}$ and obtain $U_{a}|u_{n\mathbf{k}}\rangle^{*}=\sum_{m}B_{n,m}(\mathbf{k})|u_{mD_{a}\mathbf{k}}\rangle.$
Then the non-Abelian connections are related by a non-Abelian transformation
\begin{align}
A_{\alpha}^{nn^{\prime}}(\mathbf{k})= & -i\sum_{m}B_{n,m}(\mathbf{k})\partial_{\alpha}B_{m,n^{\prime}}^{\dagger}(\mathbf{k}) -J_{\alpha\alpha^{\prime}}^{a}\sum_{m,m^{\prime}}B_{n,m}(\mathbf{k})[A_{\alpha^{\prime}}^{T}]^{mm^{\prime}}(D_{a}\mathbf{k})B_{m^{\prime},n^{\prime}}^{\dagger}(\mathbf{k})\label{eq:1}
\end{align}
where $\alpha,\alpha^{\prime}=x,y$ and $J_{\alpha\alpha^{\prime}}^{a}=\frac{\partial(D_{a}\mathbf{k})_{\alpha^{\prime}}}{\partial\mathbf{k}_{\alpha}}$
denotes element of the Jacobian matrix.

Now we consider multiple closed loops $\mathrm{C}_{i}$ , where $i=1,2,..$,
in momentum space which are invariant under the anti-unitary symmetry.
These loops can be viewed as a subspace of the whole Brillouin zone.
The sum of the line integrals of the Berry connection $\mathrm{Tr}[A_{\alpha}(\mathbf{k})]$
over these closed loops is given by
\begin{align}
\sum_{i}\ointctrclockwiseop_{C_{i}}\mathrm{Tr}(A_{\alpha}(\mathbf{k}))dl_{\alpha}^{i}= \sum_{i}\ointctrclockwiseop_{C_{i}}-i\mathrm{Tr}(B(\mathbf{k})\partial_{\alpha}B^{\dagger}(\mathbf{k}))dl_{\alpha}^{i}
  -\sum_{i}\ointctrclockwiseop_{D_{a}C_{i}}\mathrm{Tr}(A_{\alpha^{\prime}}^{T}(D_{a}\mathbf{k}))d(D_{a}\boldsymbol{l}^{i})_{\alpha^{\prime}}
\end{align}
where $d\boldsymbol{l}^{i}$ are the differential vector element along
the path $C_{i}$ and we have used $J_{\alpha\alpha^{\prime}}^{a}dl_{\alpha}^{i}=d(D_{a}\boldsymbol{l}^{i})_{\alpha^{\prime}}$
in the last line. Finally, for an anti-unitary symmetry with $\mathrm{Det}[J^{a}]=+1$,
the direction of integration along the path is preserved after the
symmetry transformation (Upper panel in Fig. 1). Consequently,
we obtain:
\begin{equation}
\sum_{i}\ointctrclockwiseop_{D_{a}C_{i}}\mathrm{Tr}[A_{\alpha^{\prime}}^{T}(D_{a}\mathbf{k})]d(D_{a}\boldsymbol{l}^{i})_{\alpha^{\prime}}=\sum_{i}\ointctrclockwiseop_{C_{i}}\mathrm{Tr}[A_{\alpha}(\mathbf{k})]dl_{\alpha}^{i}.
\end{equation}
Thus, the loop integral of the Berry connection is quantized to an
half-integer multiple of $2\pi$
\[
2\ointctrclockwiseop_{C}\mathrm{Tr}A_{\alpha}(\mathbf{k})dl_{\alpha}=-i\ointctrclockwiseop_{C}\mathrm{Tr}[B(\mathbf{k})\partial_{\alpha}B^{\dagger}(\mathbf{k})]dl_{\alpha}.
\]
The sewing matrix $B(\mathbf{k})$ associated with the anti-unitary
transformation $S_{a}$ defines a continuous map from one dimensional
circle(s) $C$ onto the unitary group $U(n)$, which is classified
by the first homotopy group $\pi_{1}(U(n))\cong\mathbb{Z}$. Here,
$n$ is the degree of the band degeneracy. Thus, the integral on the
right hand side represents the winding number of the homotopic maps
which leads to an integer $\nu$ multiple of $2\pi$ . Then, by combining
with Eq. (\ref{eq:Berry_connection}), we prove that the Hall conductivity
is equal to $\frac{\nu}{2}\frac{e^{2}}{h}$ as in Eq. (\ref{eq:quantum-Hall}),
provided that the states on the Fermi surface possess an anti-unitary
symmetry with $\mathrm{Det}[J^{a}]=+1$. It is noted that the line
integral of the Berry connection for the Hall conductivity is limited
along the Fermi surface. Thus we do not require the whole system is
invariant under the symmetry. Instead the local symmetry along the
circle of the Fermi surface is sufficient to impose the constraint
for the Hall conductance.

Furthermore, confirming that the Berry connections are related by
a non-Abelian gauge transformation, one verifies that Berry curvature
satisfies the condition
\[
\Omega_{\alpha\beta}(\mathbf{k})=-J_{\alpha\alpha^{\prime}}^{a}J_{\beta\beta^{\prime}}^{a}B(\mathbf{k})\Omega_{\alpha^{\prime}\beta^{\prime}}^{T}(D_{a}\mathbf{k})B^{\dagger}(\mathbf{k}).
\]
If the whole system is invariant under the symmetry $S_{a}$, integration
over the Brillouin zone leads to the expression
\begin{align*}
\int d^{2}\mathbf{k}\epsilon_{\alpha\beta}\mathrm{Tr}[\Omega_{\alpha\beta}(\mathbf{k})] & =-\int d^{2}\mathbf{k}\mathrm{Det}[J^{a}]\epsilon_{\alpha^{\prime}\beta^{\prime}}\mathrm{Tr}[\Omega_{\alpha^{\prime}\beta^{\prime}}(D_{a}\mathbf{k})]
\end{align*}
where we have utilized the properties $\mathrm{Tr}[M^{T}]=\mathrm{Tr}[M]$
for any matrix $M$, and $\epsilon_{\alpha\beta}J_{\alpha\alpha^{\prime}}^{a}J_{\beta\beta^{\prime}}^{a}=\epsilon_{\alpha^{\prime}\beta^{\prime}}Det[J^{a}]$.
Additionally, the transformation of the integration measure $\int d^{2}\mathbf{k}\mathrm{Det}[J^{a}]=\mathrm{sgn}(\mathrm{Det}[J^{a}])\int d^{2}(D_{a}\mathbf{k})$
yields
\begin{equation}
(1+\mathrm{sgn}[\mathrm{Det}(J^{a})])\int d^{2}\mathbf{k}\mathrm{Tr}[\Omega_{xy}(\mathbf{k})]=0.\label{eq:bulk-integral}
\end{equation}
Thus, a significant constraint is placed on the Hall conductivity,
$\sigma_{xy}=0$ when $\mathrm{Det}(J^{a})=+1$. Since it is an areal
integral of the Berry curvature over the occupied states, it is required
that the whole system is invariant under the symmetry.

Therefore, to have a nonzero $\nu$ in Eq. (\ref{eq:quantum-Hall}),
there are two conditions: (i) there exist the local symmetry along
the circle(s) of the Fermi surface; (ii) the symmetry of the whole
system must be broken.

\subparagraph*{The unitary symmetry}

For a unitary symmetry, the symmetric Bloch Hamiltonian has the relation
$S_{u}h(\mathbf{k})S_{u}^{-1}=h(D_{u}\mathbf{k})$ where $D_{u}\mathbf{k}$
is the transformed wave vector $\mathbf{k}$ under $S_{u}.$ For any
eigenstate $|u_{n\mathbf{k}}\rangle$, we have $h(D_{u}\mathbf{k})S_{u}|u_{n\mathbf{k}}\rangle=S_{u}h(\mathbf{k})|u_{n\mathbf{k}}\rangle=\varepsilon_{n\mathbf{k}}S_{u}|u_{n\mathbf{k}}\rangle.$
In this situation, we can expand $|u_{n\mathbf{k}}\rangle=\sum_{m}C_{n,m}(\mathbf{k})S_{u}^{\dagger}|u_{mD_{u}\mathbf{k}}\rangle$
where $C_{n,m}(\mathbf{k})=\langle D_{u}\mathbf{k},m|S_{u}|\mathbf{k},n\rangle$
are the sewing matrix for the unitary transformation. Then, the non-Abelian
Berry connection has the following property:
\begin{align*}
A_{\alpha}^{nn^{\prime}}(\mathbf{k})= & -i\sum_{m}C_{n,m}^{*}(\mathbf{k})\partial_{\alpha}C_{m,n^{\prime}}^{T}(\mathbf{k})\\
 & +J_{\alpha\alpha^{\prime}}^{u}\sum_{m,m^{\prime}}C_{n,m}^{*}(\mathbf{k})A_{\alpha^{\prime}}^{mm^{\prime}}(D_{u}\mathbf{k})C_{m^{\prime},n^{\prime}}^{T}(\mathbf{k})
\end{align*}
with $J_{\alpha\alpha^{\prime}}^{u}=\frac{\partial(D_{u}\mathbf{k})_{\alpha^{\prime}}}{\partial\mathbf{k}_{\alpha}}$.

Now we consider multiple closed loops $\mathrm{C}_{i}$ , where $i=1,2,..$,
in momentum space which is invariant under the unitary symmetry. Then
the Berry connection integral over these loops is
\begin{align*}
\sum_{i}\ointctrclockwiseop_{C_{i}}\mathrm{Tr}A_{\alpha}(\mathbf{k})dl_{\alpha}^{i}= & \sum_{i}\ointctrclockwiseop_{C_{i}}-i\mathrm{Tr}[C^{*}(\mathbf{k})\partial_{\alpha}C^{T}(\mathbf{k})]dl_{\alpha}^{i}\\
 & +\sum\ointctrclockwiseop_{D_{u}C_{i}}\mathrm{Tr}[A_{\alpha^{\prime}}(D_{u}\mathbf{k})]d(D_{u}\boldsymbol{l}^{i})_{\alpha^{\prime}}.
\end{align*}
For unitary symmetries with $\mathrm{Det}[J^{u}]=-1$, the direction
of integration along the path is reversed after the symmetry transformation,
as shown in the bottom panel in Fig. 1, which leads
to the following relationship
\begin{equation}
\sum_{i}\ointctrclockwiseop_{D_{a}C_{i}}\mathrm{Tr}[A_{\alpha^{\prime}}^{T}(D_{a}\mathbf{k})]d(D_{a}\boldsymbol{l}^{i})_{\alpha^{\prime}}=-\sum_{i}\ointctrclockwiseop_{C_{i}}\mathrm{Tr}[A_{\alpha}(\mathbf{k})]dl_{\alpha}^{i}.
\end{equation}
Thus, the integral of the Berry connection over the symmetry-preserving
closed loop is quantized as a half-integer multiplied by $2\pi$,
\[
2\sum_{i}\ointctrclockwiseop_{C_{i}}\mathrm{Tr}A_{\alpha}(\mathbf{k})dl_{\alpha}^{i}=-i\sum_{i}\ointctrclockwiseop_{C_{i}}\mathrm{Tr}[C^{*}(\mathbf{k})\partial_{\alpha}C^{T}(\mathbf{k})]dl_{\alpha}^{i}.
\]
Then, by incorporating Eq. (\ref{eq:Berry_connection}), we discover
that the Hall conductivity becomes half-quantized in units of $e^{2}/h$,
assuming that the states on the Fermi surface exhibit unitary symmetry
with $\mathrm{Det}[J^{u}]=-1$.

Similarly, the non-Abelian Berry curvature satisfies
\[
\Omega_{\alpha\beta}(\mathbf{k})=J_{\alpha\alpha^{\prime}}^{u}J_{\beta\beta^{\prime}}^{u}C^{*}(\mathbf{k})\Omega_{\alpha^{\prime}\beta^{\prime}}(D_{u}\mathbf{k})C^{T}(\mathbf{k}).
\]
A unitary symmetry operation with representation $S_{u}$ imposes
a constraint on the conductivity tensors $\sigma_{\alpha\beta}$ as
$\sigma_{\alpha\beta}=J_{\alpha\alpha^{\prime}}^{u}J_{\beta\beta^{\prime}}^{u}\sigma_{\alpha^{\prime}\beta^{\prime}}$.
A similar proof analogous to that used for anti-unitary symmetry can
be executed to demonstrate that: $(1-\mathrm{sgn}[\mathrm{Det}(J^{a})])\int\frac{d^{2}\mathbf{k}}{2\pi}\mathrm{Tr}[\Omega_{xy}(\mathbf{k})]=0$.
When $\mathrm{Det}(J^{u})=-1$, the Hall conductivity vanishes $\sigma_{xy}=0$.

\paragraph*{Local symmetry via global symmetry breaking}

Equations  (\ref{eq:Berry_connection}) and (\ref{eq:multiband_berry_connection})
indicate that the calculation of the anomalous Hall conductivity in
metals can be reduced to an evaluation at the Fermi surface by means
of Stokes's theorem. This formula prompts us to explore the concept
of local symmetry. Local symmetry is characterized as a symmetry that
applies exclusively only to the states at and near the Fermi surface,
rather than to the entire state space. The concept has significant
consequences for metallic systems with a Fermi surface, although the
whole system breaks the symmetry to have a nonzero Hall conductivity.
The proposed local-symmetry, where a symmetry operation applies only
to a subspace of the whole Brillouin zone and not the entire zone,
represents a deviation from traditional symmetries in physics that
generally apply to the whole system under consideration. Owing to
the fact that the line integral of the Berry connection is limited
to the regions of the Brillouin zone where this local symmetry is
present, the Hall conductivity may show a plateau as the chemical
potential varies within the region. The presence of the local symmetry
ensures that the half-quantized Hall conductivity is immune to weak
disorder or interactions, thereby stabilizing the quantized Hall metallic
phase, instead of it being a transient critical point between two
insulating phase.

\subparagraph*{Local symmetry in accidental cases}

We first present two examples to illustrate how the presence of local
symmetry can lead to the half-quantization of the Hall conductivity
in two two-band models. The first example is the Bernevig-Hughes-Zhang
model for a Chern insulator. The $k\cdot p$ Hamiltonian is expressed
as $H_{\mathrm{BHZ}}=\hbar v\mathbf{k}\cdot\boldsymbol{\sigma}+(\Delta-b\hbar^{2}\mathbf{k}^{2})\sigma_{z}$
with $\mathrm{sgn}(\Delta b)>0$ \cite{shen2017topological}. Its
Hall conductivity as a function of the chemical potential is given
by $\sigma_{H}=\frac{e^{2}}{2h}[\frac{\Delta-b\hbar^{2}k_{F}^{2}}{\mu}+\mathrm{sgn}(b)]$
where $k_{F}$ as the Fermi wave vector. As the mass term goes to
zero, $\Delta-b\hbar^{2}k_{F}^{2}=0$, the time-reversal symmetry
or the parity symmetry along the Fermi surface loop is restored, and
the Hall conductivity becomes a half-integer value of $\frac{e^{2}}{2h}\mathrm{sgn}(b)$.
However, the total system still breaks the time-reversal symmetry.
This example illustrates explicitly that the local symmetry can be
realized while the whole system breaks the symmetry, and can be regarded
as a typical example of local symmetry. Recent observed one-half anomalous
Hall effect in twisted systems is possibly relevant to this mechanism
\cite{park2023observation,lu2024fractional}.

The second example is the massless Wilson fermion by setting $\Delta=0$.
It breaks the time reversal symmetry in the presence of the $b\hbar^{2}\mathbf{k}^{2}$
term. In the low-energy limit, the system can be regarded approximately
as by linear Dirac fermions as the the symmetry broken term almost
vanishes. The parity symmetry is restored when the chemical potential
approaches the Dirac point. Consequently, the Hall conductivity is
reduced to $\frac{e^{2}}{2h}\mathrm{sgn}(b)$. This is also an example
of local symmetry near the Dirac point \cite{fu2022quantum}.

This argument can be generalized to the cases where the local symmetry
is approximately preserved. For instance, consider a two-dimensional
free electron gas with the Rashba spin-orbit coupling subjected to
a Zeeman effect: $H_{\mathrm{R}}=\frac{\hbar^{2}|\mathbf{k}|^{2}}{2m}+\lambda\mathbf{k}\cdot\boldsymbol{\sigma}+\frac{\mu_{B}g}{2}B_{z}\sigma_{z}$
where $m$ is the effective mass, $\lambda$ is the linear Rashba
coupling\cite{onoda2006intrinsic,xiao2010berry,sun2022possible}.
When the chemical potential is located within the Zeeman-induced energy
splitting, the Fermi surface is a single loop. The Hall conductivity
is given by: $\frac{e^{2}}{2h}[\frac{\mu_{B}gB_{z}/2}{\mu-\hbar^{2}k_{F}^{2}/2m}-\mathrm{sgn}(B_{z})]$.
In the case where $\mu_{B}gB_{z}\ll\frac{2m\lambda^{2}}{\hbar^{2}}$the
symmetry-breaking term is negligibly small, and the Hall conductivity
approximately equals to $-\mathrm{sgn}(B_{z})\frac{e^{2}}{2h}$.

\subparagraph*{Local symmetries in the surface states of three-dimensional topological
insulators}

In the scenarios described above, the condition to exhibit a half-integer
Hall conductivity is typically regarded as an accident occurrence,
often necessitating fine-tuning. We now turn to thin films of three-dimensional
topological (crystalline) insulator systems. These films host gapless
surface states at terminations due to the bulk's inherent topological
characteristics \cite{fu2007topological,moore2007topological,fu2007topologicalIS,fu2011topological,hsieh2012topological,ando2015topological}.
Purely 2D systems typically suffer from a fermion-doubling problem
\cite{ahn2019failure} that a single Kramers degeneracy in momentum
space must always have a counterpart elsewhere in the Brillouin zone.
Unlike the purely 2D systems, films of three-dimensional topological
(crystalline) insulators allow Kramers pairs to exist isolated on
a single two-dimensional (2D) surface \cite{moore2007topological,fu2007topological}.
The surface states of topological insulator are localized near the
two surfaces and these pairs are connected across the 3D bulk, with
their counterparts on the opposite surface, respectively.

Understanding crystal symmetry and its manifestation in the projected
surface Brillouin zone is crucial in studying surface phenomena in
three-dimensional topological systems \cite{wieder2018wallpaper,fang2019new}.
With the surface termination, the three-dimensional symmetry of the
bulk crystal is reduced, affecting the surface states and its electronic
structure. All 2D nonmagnetic surfaces adhere to one of the 17 wallpaper
groups. The spatial wallpaper group symmetries are limited to those
3D space group symmetries that preserve the surface normal vector,
including rotations around that vector, in-plane lattice translations,
mirror reflections, and glide reflections.The surface Brillouin zone
is derived by projecting the three-dimensional Brillouin zone onto
a plane aligned with the surface orientation. This projection typically
results in a two-dimensional representation of the 3D Brillouin zone
with some inherited symmetries, though generally with reduced symmetry
due to the surface's lower symmetry relative to the bulk. The center
of the surface Brillouin zone ($\bar{\Gamma}$ point) typically maintains
the full symmetry of the wallpaper group. The other high-symmetry
points in the surface Brillouin zone (like the $\bar{X}$,$\bar{M}$
points) do not exhibit the full symmetry of the wallpaper group but
have a subgroup of the full group\textquoteright s symmetry.

We illustrate the effects of local symmetry on the line integral of
the Berry connection over the Fermi surface using time reversal symmetry
($T$) and vertical mirror symmetry ($\sigma_{v}$). In a strong topological
insulator film of $\mathrm{Bi_{2}Se_{3}}$, the 2D surface for a [001]
termination adheres to wallpaper group $p3m1$, and the surface Brillouin
zone features a single Dirac cone at $\bar{\Gamma}$, leading to a
single ring structure in the Fermi loop as illustrated in Fig. 2(a).
Both symmetry transformations map the integration loop over the Fermi
surface back onto itself; however, vertical mirror symmetry ($\sigma_{v}$)
reverses the orientations, while time-reversal symmetry ($T$) leaves
the orientations unchanged. In topological crystalline insulator film
of $\mathrm{SnTe}$, the 2D surface for a [001] termination belongs
to wallpaper group $p4m$, and the surface Brillouin zone contains
four Dirac cones, resulting in multiple-ring structure in the Fermi
loop as illustrated in Fig. 2(b). The neighboring Dirac cones located
near the $\bar{X}$ ($\bar{Y}$) points are time-reversal counterparts,
and under time-reversal transformation, the integration loop over
one Dirac cone is mapped to the other without altering its orientation.
Meanwhile, for a mirror symmetry transformation (along the x-axis),
the integration loop over one pair of Dirac cones remains unchanged,
while the other pair is interchanged; more importantly, the orientation
of the integration is reversed.

In three-dimensional topological (crystalline) insulator films, the
seven symmetries can be categorized into two types: (i)$\sigma_{v}$,$T$,
and $C_{nz}T$ which map one surface onto itself; and (ii) $C_{2x}$,
$IT$, $S_{nz}T$, and $\sigma_{h}T$, which map one surface to the
opposite surface. These distinctions have significant implications
for the constraints on Hall conductivity. To illustrate the impact
of these local-symmetry constraints, we use a $\mathrm{Bi_{2}Se_{3}}$
film as an example. The prime $\mathrm{Bi_{2}Se_{3}}$ film terminated
with $[001]$ surfaces, exhibits symmetries including $I$,$C_{2x}$,
$C_{3z}$,$T$, and their combinations, such as $\sigma_{v}=C_{2x}I$,
$S_{3z}=C_{3z}\sigma_{h}$, $IT$, $C_{2x}T$, $C_{3z}T$, $\sigma_{v}T$
and $S_{3z}T$. Consequently, the Hall conductivity of this system
is zero. If an additional horizontal mirror symmetry $\sigma_{h}$
is present (such as in the {[}100{]} film of $\mathrm{Bi_{2}Se_{3}}$),
the Dirac cones at the two terminations can be combined to form one
mirror-even and one mirror-odd Dirac cone. In each mirror eigensector,
a single Dirac cone spans the entire 2D Brillouin zone, leading to
the manifestation of a half-quantized Hall effect, albeit with opposite
signs for each sector. The difference between these sectors results
in a quantized Hall conductance of $e^{2}/h$ a phenomenon we refer
to as the half quantum mirror Hall effect\cite{fu2024half}. This
effect contributes to the emergence of a topologically metallic state
of matter that preserves time-reversal invariance.

\subparagraph*{Locality of surface states and survival of local symmetry by surface
magnetic doping}

 Because
of the localized nature of these surface states, the symmetry broken
perturbation on one surface has a negligible effect on the states
at the opposing surface if the two surfaces are separated far away.
By selectively broken symmetries to open an energy gap in the surface
state on one side\cite{zhang2017anomalous,lu2021half,mogi2022experimental},
while ensuring that the Fermi energy intersects the gapless surface
state on the opposing side, the local symmetry can survive. Thus,
this provides a possible and feasible platform for realizing quantized
Hall metal states protected by local symmetry.

Firstly, we consider the introduction of a magnetic exchange field
in the middle of the film so that both surface states remain gapless,
as depicted diagrammatically in the left panel of Fig. 2(c). As demonstrated
in Ref. \cite{bai2023metallic}, when the exchange field exceeds
a critical strength, the system can exhibit nonzero Hall conductivity.
Since the exchange field in the middle layers primarily affects the
high-energy states and has minimal impact on the gapless surface states,
the system retains local $\sigma_{v}$,$C_{2x}$,$T$, $C_{3z}T$,
$IT$, and $S_{3z}T$ symmetries. Consequently, when the chemical
potential only intersects the surface states, the Hall conductivity
exhibits a half-quantized plateau with $|\nu|=2$. Next, we consider the addition
of an exchange field on one surface to open the energy gap of the
surface states, as illustrated diagrammatically in the middle panel
of Fig. 2(c). Since $\sigma_{v}$, $T$, $C_{3z}T$ only map the surface
states onto themselves, the system retains the local symmetries of
these three types when chemical potential only intersects the gapless
surface state. In this situation, the Hall conductivity becomes quantized
with $|\nu|=1$, with its sign determined by the
orientation of exchange field\cite{zou2022half,zou2023half}. Finally,
we consider an axion state where an out-of-plane Zeeman term with
opposite directions is applied to both surfaces\cite{mogi2017magnetic,xiao2018realization,zhang2019topological,chen2021using,chen2023side},
as shown in the right panel of Fig. 2(c). If the amplitudes of these
Zeeman terms are identical, the local $C_{2x}$, $IT$, and $S_{3z}T$
symmetries, which map the top surface to the bottom one, are preserved.
Therefore, when chemical potential sweeps over the surface states,
the Hall conductivity remains half-quantized even though the surface
states are gapped. Similarly, the local $\sigma_{h}T$ symmetry can
protect the half-quantization of Hall conductivity in a $\mathrm{Bi_{2}Se_{3}}$
film with {[}100{]} terminations.

\paragraph*{Classification of the single gapless Dirac cone with symmetry $C_{nz}$ and $T$}

In this section, we explore the constraints imposed on the form of
surface states at high-symmetry points within the surface Brillouin
zone, arising from the presence of time-reversal symmetry  ($T$) and rotational symmetry ($C_{nz}$). We then explore
how the generalized no-go theorem, which typically prevents the existence
of a single gapless Dirac cone with a nonzero topological number throughout
the entire Brillouin zone in lattice systems preserving certain symmetries,
can be circumvented by introducing symmetry-breaking elements.

\subparagraph*{Characteristics of surface states under both $C_{nz}$ and $T$ symmetries}

The surface states Hamiltonian in general has the following $2\times2$
form 
\begin{equation}
H(\mathbf{k})=f(\mathbf{k})\sigma_{+}+f^{*}(\mathbf{k})\sigma_{-}+g(\mathbf{k})\sigma_{z}\label{eq:surface_state_H}
\end{equation}
where $\sigma_{\pm}=(\sigma_{x}\pm i\sigma_{y})/2$ and $f$ and $g$
are complex and real functions of $\mathbf{k}=(k_{x},k_{y})$, respectively,
due to the Hermiticity of $H$. We consider the symmetry constraints
imposed on $H(\mathbf{k})$ in the presence of time reversal symmetry
$T$ and the additional n-fold rotation symmetry $C_{n}$ with respect
to the z axis:
\begin{align}
TH(\mathbf{k})T^{-1} & =H(-\mathbf{k})\label{eq:tr_constraint}\\
C_{n}H(\mathbf{k})C_{n}^{-1} & =H(D_{n}\mathbf{k})\label{eq:Cn_constraint}
\end{align}
where $D_{n}$ represents the rotation around the $z$ axis in momentum
space. Here, $n$ is restricted to be $n=2,3,4,6$ in periodic lattice
systems. Let us first consider surface states in the vicinity of a
high symmetry point invariant under $C_{n}$ and time reversal operations,
such as $\bar{\Gamma}$ and $\bar{M}$-points. Since we are interested
in the form of gapless structures at $\mathbf{k}\simeq0$, we expand
$f$ and $g$ to leading order in $\mathbf{k}$\citep{yang2014classification},
\begin{equation}
f(k_{+},k_{-})=vk_{+}^{p}k_{-}^{q},g(k_{+},k_{-})=v^{\prime}k_{+}^{r}k_{-}^{s}+H.c.\label{eq:k_expansion_of_fg}
\end{equation}
where $v,v\prime\in\mathbb{C}$ and $p,q,r,s$ are non-negative integers,
$k_{\pm}=k_{x}\pm ik_{y}$.

If $g(\mathbf{k})=0$, the Hamiltonian for the surface states contains
only two Dirac matrices, expressed as $H(\mathbf{k})=f(\mathbf{k})\sigma_{+}+f^{*}(\mathbf{k})\sigma_{-}$.
Owing to an additional sub-lattice symmetry $\{\Gamma,H(\mathbf{k})\}=0$
with $\Gamma=-\sigma_{z}$, we can define the $\mathbb{Z}$-valued
topological number $W_{1}$ along a loop $C$\citep{chiu2014classification,chiu2016classification},
\begin{equation}
W_{1}=\frac{1}{4\pi i}\oint_{C}d\mathbf{k}\cdot\mathrm{Tr}[\Gamma H^{-1}(\mathbf{k})\nabla_{\mathbf{k}}H(\mathbf{k})].\label{eq:winding_number-1}
\end{equation}
By substituting the Hamiltonian in Eq. (\ref{eq:winding_number-1}),
the winding number reduced to $W_{1}=\frac{1}{2\pi}\oint_{C}d\mathbf{k}\cdot\nabla_{\mathbf{k}}\mathrm{Arg}[f(\mathbf{k})].$
Let $D_{i}$ be a disk enclosing an $i$th Dirac point, so the total
winding number is 
\[
W_{1}=\frac{1}{2\pi}\oint_{\cup_{i}\partial D_{i}}d\mathbf{k}\cdot\nabla_{\mathbf{k}}\mathrm{Arg}[f(\mathbf{k})]
\]
with $\partial D_{i}$ as the boundary of $D_{i}$. If $f(\mathbf{k})=v_{+}k_{+}^{n}+v_{-}k_{-}^{n}$,
the winding number $W_{1}=-n$ for $|v_{+}|>|v_{-}|$ and $W_{1}=+n$
for $|v_{+}|<|v_{-}|$. A non-vanishing integer value of $W_{1}$
implies that the gapless point in the integration loop is stable and
cannot be annihilated by itself.

(i) For spin $1/2$ system, the TRS operator $T$ can be represented as
$T=i\sigma_{y}K$ with $T^{2}=-1$, where $\sigma_{x,y,z}$ are Pauli
matrices for spin degrees of freedom and $K$ stands for complex conjugation.
The invariance of the Hamiltonian under TR in Eq. (\ref{eq:tr_constraint})
leads to the relations: $f(-\mathbf{k})=-f(\mathbf{k})$ and $g^{*}(-\mathbf{k})=-g(\mathbf{k})$.
Thus, both $f$ and $g$ are odd functions of $\mathbf{k}$. When
$D_{n}\mathbf{k}=\mathbf{k}$ is satisfied, we have $[C_{n},H(\mathbf{k})]=0$.
Using a basis in which $C_{n}$ is diagonal, $C_{n}$ can be represented
by a diagonal matrix $C_{n}=diag[\alpha_{l},\alpha_{m}],$ where $\alpha_{l}=\exp(i\frac{2\pi}{n}(l+\frac{1}{2}))$
with $l=0,1,...,n-1$ . For convenience, we can express $C_{n}$ as
$C_{n}=e^{i\pi(\frac{1+l+m}{n}+\frac{l-m}{n}\sigma_{z})}$. We assume
the commutation relation$[C_{n},T]=0$, so that the representation
of the $C_{n}$ operator is restricted to $C_{n}=e^{i\pi(\frac{2l+1}{n}\sigma_{z})}$.
From Eq. (\ref{eq:Cn_constraint}), we have
\begin{align}
e^{i2\pi(2l+1)/n}f(e^{i2\pi/n}k_{+},e^{-i2\pi/n}k_{-}) & =f(k_{+},k_{-}),\label{eq:symmetry_constraint_on_fg}\\
g(e^{i2\pi/n}k_{+},e^{-i2\pi/n}k_{-}) & =g(k_{+},k_{-}).\nonumber 
\end{align}
Substituting Eq. (\ref{eq:k_expansion_of_fg}) into Eqs. (\ref{eq:symmetry_constraint_on_fg})
yields the relations:
\begin{align*}
e^{i2\pi(2l+1+p-q)/n} & =1,e^{i2\pi(r-s)/n}=1
\end{align*}
which gives
\begin{align*}
q-p & =2l+1,\mathrm{mod}\;n,\\
r-s & =0,\mathrm{mod}\;n.
\end{align*}
For $n=2,4,6$, since $g(k_{+},k_{-})$ is odd function, we find $g(\mathbf{k})=0$.
For $n=3$, the lowest order non-vanishing term for $g$ can be expressed
as 
\[
g(k_{+},k_{-})=v^{\prime}k_{+}^{3}+v^{\prime*}k_{-}^{3}.
\]
Table 2 provides an overview of the symmetry- permitted form of $f(\mathbf{k})$
and $g(\mathbf{k})$ for small $\mathbf{k}$. In systems exhibiting
a three fold rotation symmetry, such as $\mathrm{Bi}_{2}\mathrm{Se}_{3},\mathrm{Bi}_{2}\mathrm{Te}_{3}$,
a hexagonal warping term $(v^{\prime}k_{+}^{3}+v^{\prime*}k_{-}^{3})\sigma_{z}$
is allowed by symmetry\citep{fu2009hexagonal,liu2010model}. This
leads to an energy dispersion that deviates from a circular shape,
often producing a snowflake-like pattern. However, the presence of
time reversal symmetry ensures that the line integral of Berry connection
along the loop of a constant energy for such a system remains quantized
to $\pi$, thereby upholding the half quantization of Hall conductivity
in a semi-magnetic $\mathrm{Bi}_{2}\mathrm{Se}_{3}$ film. For BHZ
model with $C_{4}$ symmetry, such a warping term is excluded by symmetry\citep{konig2007quantum,konig2008quantum}.

(ii) For spinless system, TRS can be represented by the operator $T=\sigma_{x}K$
with $T^{2}=+1$, where $\sigma_{x,y,z}$ are Pauli matrices for pseudo-spin
basis and $K$ stands for complex conjugation. The invariance of the
Hamiltonian under TR as stated in Eq. (\ref{eq:tr_constraint}) leads
to the relations: $f(-\mathbf{k})=f(\mathbf{k})$ and $g^{*}(-\mathbf{k})=-g(\mathbf{k})$.
Thus, $f$ and $g$ are even and odd functions of $\mathbf{k}$, respectively.
When $D_{n}\mathbf{k}=\mathbf{k}$ is satisfied, $[C_{n},H(\mathbf{k})]=0$.
In a basis characterized by orbital angular momentum $l$ and $m$,
$C_{n}$ can be represented by a diagonal matrix $C_{n}=diag[\alpha_{l},\alpha_{m}],$
where $\alpha_{l}=\exp(i\frac{2\pi}{n}l)$ with $l=0,1,...,n-1$.
This can be expressed as $C_{n}=e^{i\pi(\frac{l+m}{n}+\frac{l-m}{n}\sigma_{z})}$.
Assuming the commutation relation $[C_{n},T]=0$, so that the representation
of the $C_{n}$ operator is restricted to $C_{n}=e^{i\frac{2\pi l}{n}\sigma_{z}}$.
From Eq. (\ref{eq:Cn_constraint}), we have
\begin{align}
e^{i\frac{4\pi l}{n}}f(e^{i2\pi/n}k_{+},e^{-i2\pi/n}k_{-}) & =f(k_{+},k_{-}),\label{eq:symmetry_constraint_on_fg-1}\\
g(e^{i2\pi/n}k_{+},e^{-i2\pi/n}k_{-}) & =g(k_{+},k_{-})\nonumber 
\end{align}
From Eq. (\ref{eq:k_expansion_of_fg}), we have 
\begin{align*}
e^{i\frac{2\pi}{n}(2l+p-q)} & =1,e^{i\frac{2\pi}{n}(r-s)}=1
\end{align*}
resulting in the conditions:
\begin{align*}
q-p & =2l,\mathrm{mod}\;n\\
r-s & =0,\mathrm{mod}\;n.
\end{align*}
For $g(\mathbf{k}_{\shortparallel})$, the symmetry constraints are
the same as those for the spin-1/2 system.

When $\mathrm{mod}(2l,n)=0$, the leading term of $f(\mathbf{k})$
is a constant. To properly reflect the momentum dependence, $f(\mathbf{k})$
should be expanded to include higher order terms of $\mathbf{k}$.
In this case, the surface states at the high-symmetry point $\mathbf{k}$
is split into $n$ Dirac cones with linear dispersion at generic momenta
$\mathbf{k}_{0},D_{n}\mathbf{k}_{0},...,D_{n}^{n-1}\mathbf{k}_{0}$
with $f(\mathbf{k}_{0})=0$. For instance, when examining the case
where $n=2$ , we limit our consideration to the second lowest order
of $\mathbf{k}$: $f(\mathbf{k})=v+v_{+}k_{+}^{2}+v_{-}k_{-}^{2}$\citep{kobayashi2021fragile}.
Given that $f(\mathbf{k}_{0})=0$, the presence of 2-fold rotational
symmetry necessitates that $f(-\mathbf{k}_{0})=0$ as well. We can
then expand $f(\mathbf{k})$ in the vicinity of $\pm\mathbf{k}_{0}$,such
that $f_{\pm}(\mathbf{k})\simeq\pm2(v_{+}k_{0+}\delta k_{+}+v_{-}k_{0-}\delta k_{-})$.
Therefore, the winding number $W_{1}$ around the two Dirac points
is of the same sign, which leads to a total winding number expressed
as $2\mathrm{sgn}(|v_{-}|-|v_{+}|)$. When $\mathrm{mod}(2l,n)\ne0$,
the leading term of $f(\mathbf{k})$ is proportional to $|\mathbf{k}|^{2}$
for $n=4$, resulting in the surface states with quadratic dispersion
at $\mathbf{k}$, as in the Fu model\citep{fu2011topological}. This
quadratic band touching is linked with a two-fold degeneracy enforced
by 2D representations for $n=4$.

\subparagraph*{The generalized no-go theorem}

When the gapless cone, as shown in Table 2, is implemented on a lattice
while preserving time-reversal symmetry, it encounters the fermion
doubling problem. For systems with 2-fold rotational symmetry (e.g.,
including 2, 4, 6-fold symmetries), an additional chiral symmetry
for the surface states allows us to define a winding number for a
1D integral path over the Fermi loop, which is a $\mathbb{Z}$-type
invariant\citep{chiu2014classification}. However, in systems exhibiting
three-fold rotational symmetry, where chiral symmetry is absent, the
Berry phase along any closed loop with local time-reversal symmetry
preserved is quantized. This Berry phase represents a $\mathbb{Z}_{2}$-type
invariant, as it is defined only up to multiples of $2\pi$\citep{chan20163}.
A stable topological point at $\mathbf{k}$ is characterized by a
non-zero topological invariant, which can be either $\mathbb{Z}$
or $\mathbb{Z}_{2}$. A non-zero topological invariant for a 1D integral
path corresponds to a topological charge enclosed by the loop. The
generalized no-go theorem is that the sum of the topological charges
carried by the topological points across the entire Brillouin Zone
must cancel out, achieving charge neutralization\citep{nielsen1981absenceI,nielsen1981absenceII}.
For system with $C_{3}$ symmetry, when the Berry phase is $\pi$
corresponding $\mathbb{Z}_{2}=1$, an additional Dirac cone is required
to offset this charge neutralization discrepancy. For system with
2-fold rotational symmetry, any topological point with a nonzero winding
number must be paired with another topological point that has an opposite
winding number\citep{ahn2019failure,le2022generalized}.

In order to maintain a single gapless structure in Eq. (\ref{eq:surface_state_H})
at low energy scales and to avoid the fermion doubling problem, time-reversal
symmetry must be explicitly broken in the high energy regime. We consider
a Hamiltonian of the form $H(\mathbf{k})=f(\mathbf{k})\sigma_{+}+f^{*}(\mathbf{k})\sigma_{-}+m(\mathbf{k})\sigma_{z}$
with $m(\mathbf{k})=g(\mathbf{k})$ in the low energy regime and $m(\mathbf{k})=V_{z}$
in the high energy regime which breaks time-reversal symmetry explicitly\citep{zou2022half,zou2023half}.
The wave function for the valence band is given by $|u_{\mathbf{k}}^{-}\rangle=\frac{1}{\sqrt{2}}\left(\begin{array}{cc}
-\sqrt{1-\frac{m}{\epsilon}}e^{i\mathrm{Arg}[f]}, & \sqrt{1+\frac{m}{\epsilon}}\end{array}\right)^{T}$ for $V_{z}>0$ and $|u_{\mathbf{k}}^{-}\rangle=\frac{1}{\sqrt{2}}\left(\begin{array}{cc}
-\sqrt{1-\frac{m}{\epsilon}}, & \sqrt{1+\frac{m}{\epsilon}}e^{-i\mathrm{Arg}[f]}\end{array}\right)^{T}$ for $V_{z}<0$ to ensure that the wave function is singularity-free
throughout the Brillouin zone. Here, $\epsilon=\sqrt{|f|^{2}+m^{2}}$.
Subsequently, the Hall conductivity can be calculated as, 
\[
\sigma_{xy}=\frac{e^{2}}{2h}\mathrm{sgn}(V_{z})\int_{\mathrm{C}_{FS}}d\mathbf{l}\cdot\partial_{\mathbf{k}}\mathrm{Arg}[f]-\frac{e^{2}}{2h}\int_{\mathrm{C}_{FS}}d\mathbf{l}\cdot\frac{g}{\epsilon}\partial_{\mathbf{k}}\mathrm{Arg}[f]
\]
where the first term simplifies to $\frac{e^{2}}{2h}\mathrm{sgn}(V_{z})W_{1}$
and the second term vanishes due to the property that $g(\mathbf{k}_{F})=-g(-\mathbf{k}_{F})$
under time reversal symmetry. This result indicates the Hall conductivity for each type of Dirac fermion listed in Table 2, resulting from the introduction of the time-reversal symmetry-breaking term.

\paragraph*{Two Materials}

In this section, we numerically validate the discussions presented
above by examining semi-magnetic films of the strong topological insulators
$\mathrm{Bi_{2}Te_{3}}$ and $\mathrm{Bi_{2}Se_{3}}$, as well as
the topological crystalline insulator $\mathrm{SnTe}$. We show that
$\nu=1$ for $\mathrm{Bi_{2}Te_{3}}$ and $\mathrm{Bi_{2}Se_{3}}$
and $\nu=2$ or $4$ for $\mathrm{SnTe}$.

\subparagraph*{Strong Topological Insulators $\mathrm{Bi_{2}Te_{3}}$ and $\mathrm{Bi_{2}Se_{3}}$}

The strong three-dimensional (3D) topological insulators (TIs), $\mathrm{Bi_{2}Te_{3}}$ and $\mathrm{Bi_{2}Se_{3}}$, exhibit a rhombohedral crystal structure that falls within
the R3m space group (No. 166)\citep{zhang2009topological,liu2010model}.
To compute the electronic structure of a TI film, we employ a tight-binding
model encompassing four states, specifically, $|P1_{-}^{+},\pm\frac{1}{2}\rangle$and
$|P2_{+}^{-},\pm\frac{1}{2}\rangle$, with parameters determined by
fitting to electronic properties derived from first-principles calculations\citep{acosta2018tight}.
The Hamiltonian's form is significantly constrained by crystal symmetries
and time-reversal symmetry, characterized by: i) threefold rotation
symmetry $C_{3z}$ along the z axis, ii) twofold rotation symmetry
$C_{2x}$ along the $x$ axis, iii) inversion symmetry $P$, and iv)
time-reversal symmetry $T$. Fig. 3(a) illustrates the 3D Brillouin
zone and the projected surface Brillouin zones for the [001], [100],
and [010] surfaces of the rhombohedral structure, each featuring a
single Dirac cone. However, their properties vary considerably due
to different symmetries. For the [001] surface, a hexagonal warping
term allowed by $C_{3z}$ symmetry results in a non-circular, snowflake-like
energy dispersion. For the [100] surface, the warping term is precluded
by $C_{2x}$ symmetry. For the [010] surface, which possesses only
mirror symmetry $M_{x}$(depicted by the shaded plane in Fig. 3(a)),
the lowest-order momentum term allowed in $g(\mathbf{k})$ is linear.
By introducing a time-reversal symmetry-breaking Zeeman term on one
surface of the [001] and [100] terminations\citep{song2022first},
\[
H^{\prime}=\sum_{\mathbf{r}}\psi_{\mathbf{r}}^{\dagger}f(\mathbf{r})\frac{\mu_{B}}{2}\sum_{i=x,y,z}[g_{1i}B_{i}\tau_{0}\sigma_{i}+g_{2i}B_{i}\tau_{z}\sigma_{i}]\psi_{\mathbf{r}}
\]
where the creation field operator $\psi_{\mathbf{r}}^{\dagger}=(c_{\mathbf{r},+,\uparrow}^{\dagger},c_{\mathbf{r},+,\downarrow}^{\dagger},c_{\mathbf{r},-,\uparrow}^{\dagger},c_{\mathbf{r},-,\downarrow}^{\dagger})$
with $\pm$ denoting the parity of the basis and $\uparrow/\downarrow$
the spin. The function $f(\mathbf{r})$ equals 1 for several top layers
and 0 for the remaining layers. $g_{1i}\pm g_{2i}$ are the $g$-factors
for the two respective orbits. We observe in the band structure that
a gap opens on top surface but remains gapless on the bottom surface,
as shown in Fig. 3(b) and (d). The Hall conductivity, depicted in
Fig. 3(c) and (e), is quantized as $-e^{2}/2h$ when the chemical
potential only intersects the gapless Dirac cone \citep{zou2023half}.
This half-quantization is attributed to the persistence of local time-reversal
symmetry. For the [001] termination, the application of an in-plane
Zeeman field along $x$ direction which breaks the local time-reversal
symmetry but maintains mirror symmetry $M_{x}$, such that the Hall
conductivity will still be half-quantized according to our theory.

\subparagraph*{Topological Crystalline Insulator SnTe}

The compound $\mathrm{SnTe}$ is known to crystallize in the rock-salt
structure, which has the space group Fm-3m (in Hermann-Mauguin notation)
or 225 (in the International Tables for Crystallography)\citep{hsieh2012topological,ando2015topological}.
The space group Fm-3m (No. 225) is one of the cubic space groups and
part of the face-centered cubic (fcc) Bravais lattice. The generators
for the space group Fm-3m typically include: (i) Fourfold rotation
($C_{4}$) about the $[001]$ axis, (ii) Threefold rotation ($C_{3}$)
about the {[}111{]} diagonal, (iii) Inversion ($P$), (iv) Translation
by one half of the unit cell along the face diagonal ($\mathbf{t}$).
$\mathrm{SnTe}$ also exhibits time-reversal symmetry, denoted by
$T$. The combination of these symmetry operations generates the entire
set of operations for the Fm-3m space group. The tight-binding model
of $\mathrm{SnTe}$ is constructed from the Wannier functions of the
conduction and valence bands, which are primarily three $p$-orbitals
of Sn and Te atoms. The tight-binding Hamiltonian $\mathcal{H}_{\mathrm{\mathrm{SnTe}}}$
is given by\citep{hsieh2012topological}
\begin{align*}
\mathcal{H}_{\mathrm{\mathrm{SnTe}}} & =m\sum_{j}(-1)^{j}\sum_{\mathbf{r},\alpha}\mathbf{c}_{j\alpha}^{\dagger}(\mathbf{r})\cdot\mathbf{c}_{j\alpha}(\mathbf{r})\\
 & +\sum_{j,j^{\prime}}t_{jj^{\prime}}\sum_{(\mathbf{r},\mathbf{r}^{\prime}),\alpha}\mathbf{c}_{j\alpha}^{\dagger}(\mathbf{r})\cdot\hat{\mathbf{d}}_{\mathbf{r}\mathbf{r}^{\prime}}\hat{\mathbf{d}}_{\mathbf{r}\mathbf{r}^{\prime}}\cdot\mathbf{c}_{j^{\prime}\alpha}(\mathbf{r}^{\prime})+h.c.\\
 & +\sum_{j}i\lambda_{j}\sum_{\mathbf{r},\alpha,\beta}\mathbf{c}_{j\alpha}^{\dagger}(\mathbf{r})\times\mathbf{c}_{j\beta}(\mathbf{r})\cdot\boldsymbol{\sigma}_{\alpha\beta}
\end{align*}
Here $\mathbf{r}$ denotes the lattice site, while $j=1,2$ distinguishes
the $\mathrm{Sn}$ or $\mathrm{Te}$ atom. The electrons's spin is
labeled by $\alpha=\uparrow,\downarrow$. The components of the creation
operator vector $\mathbf{c}^{\dagger}$ and the annihilation operator
vector $\mathbf{c}$ correspond to the three $p$-orbitals. The parameter
$m$ represents the on-site potential difference between $\mathrm{Sn}$
and $\mathrm{Te}$. The hopping amplitude between nearest-neighbor
$\mathrm{Sn}$ and $\mathrm{Te}$ is symmetrical, $t_{12}=t_{21}$.
The parameters $t_{11}$ and $t_{22}$ quantify the next nearest-neighbor
hopping amplitudes within the same sub-lattice. The unit vector $\hat{\mathbf{d}}_{\mathbf{r}\mathbf{r}^{\prime}}=\frac{\mathbf{r}-\mathbf{r}^{\prime}}{|\mathbf{r}-\mathbf{r}^{\prime}|}$
connects sites $\mathbf{r}$ to $\mathbf{r}^{\prime}$. The term $\lambda_{1}$
and $\lambda_{2}$ represents the spin-orbit couplings for the two
types of atoms. The bulk and surface bands of the above tight-binding
Hamiltonian nicely reproduce the essential features of the first-principles
calculation. Fig. 4(a) displays the 3D Brillouin zone of the rock-salt
crystal structure. Crystals of the SnTe class possess small direct
band gaps at four symmetry-related time-reversal invariant momenta
(TRIMs), specifically at the L points (i.e., $L_{1}$, $L_{2}$, $L_{3}$
and $L_{4}$ in Fig. 4(a)). The topological character of the crystal
is defined by a mirror Chern number $N_{M}=(N_{+i}-N_{-i})/2$ associated
with the plane passing through $\Gamma$ and $L_{3},L_{4}$ points
which can be tuned via the model parameter $m$. Here, $N_{\pm i}$
denote the Chern numbers of the Bloch eigenstates with mirror eigenvalues
of $\pm i$, respectively. As illustrated in Fig. 4(b), the mirror
Chern number is plotted as a function of the parameter $m$. The value
of $N_{M}$ transitions from $-2$ to $0$ as $m$ increased, crossing
a critical value $m_{c}\simeq2.06\mathrm{eV}.$ Correspondingly, the
3D band structure along high symmetry lines exhibits a band inversion
when $m$ surpasses $m_{c}$ as demonstrated in Fig. 4(c). For a nonzero
mirror Chern number case, depending on the surface orientation {[}(001)
or (110){]}, there are two types of TCI surface states with qualitatively
different electronic properties as schematically shown in Fig. 4(a).
For the (001) surface termination, the existence of two gapless Dirac
cones along $\bar{\Gamma}-\bar{X}$($\bar{\Gamma}-\bar{X}^{\prime}$)
is ensured by the nonzero mirror Chern number associated with the
$\Gamma L_{3}L_{4}$($\Gamma L_{1}L_{2}$) plane, as highlighted by
the brown-colored surface in Fig. 4(a). Conversely, on (110) surface
termination, there are only two gapless Dirac cones along along $\bar{\Gamma}-\bar{Y}$.

Considering a slab geometry with two distinct terminations, we introduced
a time-reversal symmetry-breaking term on one termination, described
by the Hamiltonian

\[
\mathcal{H}^{\prime}=\Delta\sum_{j}(-1)^{j}\sum_{\mathbf{r},\alpha,\beta}f(\mathbf{r})(\boldsymbol{\sigma}\cdot\mathbf{g})_{\alpha\beta}\mathbf{c}_{j\alpha}^{\dagger}(\mathbf{r})\cdot\mathbf{c}_{j\beta}(\mathbf{r}),
\]
where $\mathbf{g}=(0,0,1)$ or $\frac{1}{\sqrt{2}}(1,1,0)$ corresponds
to the two different terminations, $\Delta$ represents the staggered
Zeeman field, and $f(\mathbf{r})$ equals $1$ for several bottom
layers and $0$ for the remaining layers. Figures 4(d) and (f) illustrate
that while the Dirac cones on the bottom surface exhibit a band gap
(indicated by blue), the Dirac cones on the top surface remain gapless
which retains the local time-reversal symmetry (signified by red).
Consequently, if the chemical potential intersects only these gapless
surface states, the Hall conductivity is quantized at $+e^{2}/h$
for $[110]$ termination and $-2e^{2}/h$ for $[001]$ terminations
depicted in Figs. 4(e) and (g).

\paragraph*{Extrinsic anomalous Hall conductivity}

As the chemical potential is located within the band, $\sigma_{xy}^{\mathrm{I}}$
may also contribute. We now turn our attention to the extrinsic contributions
to the anomalous Hall effect, focusing on mechanisms such as skew
scattering and side-jump\citep{sinitsyn2007anomalous,yang2011scattering,lu2013extrinsic}.
Given that the extrinsic contribution primarily arises from states
on the Fermi surface, our analysis will be centered on the surface
state Hamiltonian in Eq. (\ref{eq:surface_state_H}). The side-jump
contribution (upper panel in Fig. 5) to the Hall conductivity can
be expressed as follows:
\begin{align*}
\sigma_{xy}^{sj(a)} & =\frac{e^{2}\hbar}{\pi}\int_{\mathbf{k},\mathbf{k}^{\prime}}\mathrm{Re}[\langle U_{\mathbf{k}\mathbf{k}^{\prime}}^{++}U_{\mathbf{k}^{\prime}\mathbf{k}}^{-+}\rangle\\
 & \times G_{\mathbf{k}+}^{R}\gamma_{x}^{++}(\mathbf{k})G_{\mathbf{k}+}^{A}G_{\mathbf{k}^{\prime}+}^{R}v_{y}^{+-}(\mathbf{k}^{\prime})G_{\mathbf{k}^{\prime}-}^{A}],
\end{align*}
\begin{align*}
\sigma_{xy}^{sj(b)} & =\frac{e^{2}\hbar}{\pi}\int_{\mathbf{k},\mathbf{k}^{\prime}}\mathrm{Re}[\langle U_{\mathbf{k}\mathbf{k}^{\prime}}^{++}U_{\mathbf{k}^{\prime}\mathbf{k}}^{+-}\rangle\\
 & \times\gamma_{x}^{++}(\mathbf{k})G_{\mathbf{k}+}^{R}G_{\mathbf{k}^{\prime}+}^{R}G_{\mathbf{k},-}^{R}v_{y}^{-+}(\mathbf{k})G_{\mathbf{k},+}^{A}].
\end{align*}
The integral over the momentum space is concisely denoted by$\int_{\mathbf{k}}=\int\frac{d^{2}\mathbf{k}}{(2\pi)^{2}}$.
$\langle U_{\mathbf{k}\mathbf{k}^{\prime}}^{ss^{\prime}}U_{\mathbf{k}^{\prime}\mathbf{k}}^{r^{\prime}r}\rangle$
represents the disorder-averaged scattering potential matrix elements
where $r,s,r^{\prime},s^{\prime}$ are band indices. $G_{\mathbf{k}s}^{R/A}=(\mu-\epsilon_{\mathbf{k}}^s\pm i\eta)^{-1}$
denotes the retarded/advanced Green's function for band $s$, where $\epsilon_{\mathbf{k}}^s$ denotes the corresponding energy and $\eta$ is the broadening induced by disorder. $v_{y}^{ss^{\prime}}(\mathbf{k})$
and $\gamma_{\alpha}^{ss^{\prime}}(\mathbf{k})$ are the velocity
matrix element and remonetized velocity matrix element with vertex
correction, respectively.

Considering a system with time reversal symmetry, we have $\epsilon_{\mathbf{k}}^{s}=\epsilon_{-\mathbf{k}}^{s}$
with $s=\pm$, $v_{\alpha}^{ss^{\prime}}(\mathbf{k})=-v_{\alpha}^{s^{\prime}s}(-\mathbf{k})$
and $U_{\mathbf{k}\mathbf{k}^{\prime}}^{ss^{\prime}}=U_{-\mathbf{k}^{\prime},-\mathbf{k}}^{s^{\prime}s}$.
These symmetries lead to the following relations for the side-jump
contributions to the Hall conductivity:
\begin{align*}
\sigma_{xy}^{sj(a)} & =\frac{e^{2}\hbar}{\pi}\int_{\mathbf{k},\mathbf{k}^{\prime}}\mathrm{Re}[\langle U_{\mathbf{k}\mathbf{k}^{\prime}}^{++}U_{\mathbf{k}^{\prime}\mathbf{k}}^{-+}\rangle\gamma_{x}^{++}(\mathbf{k})v_{y}^{+-}(\mathbf{k}^{\prime})]\\
 & \times G_{\mathbf{k}+}^{R}G_{\mathbf{k}+}^{A}\mathrm{Re}[G_{\mathbf{k}^{\prime}+}^{R}G_{\mathbf{k}^{\prime}-}^{A}],
\end{align*}
\begin{align*}
\sigma_{xy}^{sj(b)} & =\frac{e^{2}\hbar}{\pi}\int_{\mathbf{k},\mathbf{k}^{\prime}}\mathrm{Re}[\langle U_{\mathbf{k}\mathbf{k}^{\prime}}^{++}U_{\mathbf{k}^{\prime}\mathbf{k}}^{+-}\rangle\gamma_{x}^{++}(\mathbf{k})v_{y}^{-+}(\mathbf{k})]\\
 & \times G_{\mathbf{k}+}^{R}G_{\mathbf{k},+}^{A}\mathrm{Re}[G_{\mathbf{k}^{\prime}+}^{R}G_{\mathbf{k},-}^{R}].
\end{align*}
For a generic surface state Hamiltonian in Eq. (\ref{eq:surface_state_H}),
the term involving scattering potentials and velocities can be expressed
as $\mathrm{Re}[\langle U_{\mathbf{k}\mathbf{k}^{\prime}}^{++}U_{\mathbf{k}^{\prime}\mathbf{k}}^{-+}\rangle\gamma_{x}^{++}(\mathbf{k})v_{y}^{+-}(\mathbf{k}^{\prime})]  \propto  \sin(\theta)\sin(\phi)\{\cos(\theta^{\prime})\cos(\phi^{\prime})[\sin(\theta)\cos(\theta^{\prime})\cos(\phi-\phi^{\prime})-\cos(\theta)\sin(\theta^{\prime})]-\sin(\theta)\sin(\phi^{\prime})\sin(\phi-\phi^{\prime})\}$
with
$\cos\theta=\frac{g(\mathbf{k})}{\mu}$ and $\phi=\mathrm{Arg}[f(\mathbf{k})]$.
For systems possessing 3-fold, 4-fold or 6-fold rotational symmetry,
the side-jump contributions $\sigma_{xy}^{sj(a)}=\sigma_{xy}^{sj(b)}=0$
are zero after summing over points related by symmetry, such as $\mathbf{k},D_{n}\mathbf{k},...,D_{n}^{n-1}\mathbf{k}$.
In system with $2$-fold rotational symmetry, these contributions
cannot be directly ruled out by symmetry analysis alone. However,
considering the nature of $\mathrm{Re}[G_{\mathbf{k}^{\prime}+}^{R}G_{\mathbf{k}^{\prime}-}^{A}]\simeq\frac{\mu-\epsilon_{\mathbf{k}^{\prime}}}{2\mu\left(\eta^{2}+(\epsilon_{\mathbf{k}^{\prime}}-\mu)^{2}\right)}$,
it becomes apparent that states slightly above and below the Fermi
surface dominate the contribution, but with opposite signs. Therefore,
in this situation, the side-jump contribution $\sigma_{xy}^{sj}$
is vanishingly small.

Skew scattering contributions arise from the asymmetric part of the
scattering rates in higher-order scattering processes. The leading
contribution is associated with the third-order disorder correlation
and inversely depends on the impurity concentration $n$ shown diagrammatically
as lower panel in Fig. 5. The skew scattering conductivity $\sigma_{xy}^{sk}$
can be written as:
\begin{align*}
\sigma_{xy}^{sk} & =\frac{e^{2}\hbar}{\pi}\int_{\mathbf{k},\mathbf{k}^{\prime},\mathbf{k}^{\prime\prime}}\mathrm{Re}[\langle U_{\mathbf{k}\mathbf{k}^{\prime}}^{++}U_{\mathbf{k}^{\prime}\mathbf{k}^{\prime\prime}}^{++}U_{\mathbf{k}^{\prime\prime}\mathbf{k}}^{++}\rangle\\
 & \times G_{\mathbf{k}+}^{A}\gamma_{x}^{++}(\mathbf{k})G_{\mathbf{k},+}^{R}G_{\mathbf{k}^{\prime}+}^{R}\gamma_{y}^{++}(\mathbf{k}^{\prime})G_{\mathbf{k}^{\prime}+}^{A}G_{\mathbf{k}^{\prime\prime}+}^{A}].
\end{align*}
By using time-reversal symmetry, the expression simplifies further
as:
\begin{align*}
\sigma_{xy}^{sk} & =\frac{e^{2}\hbar}{\pi}\int_{\mathbf{k},\mathbf{k}^{\prime},\mathbf{k}^{\prime\prime}}\mathrm{Re}[\langle U_{\mathbf{k}\mathbf{k}^{\prime}}^{++}U_{\mathbf{k}^{\prime}\mathbf{k}^{\prime\prime}}^{++}U_{\mathbf{k}^{\prime\prime}\mathbf{k}}^{++}\rangle\gamma_{x}^{++}(\mathbf{k})\gamma_{y}^{++}(\mathbf{k}^{\prime})]\\
 & \times G_{\mathbf{k}+}^{A}G_{\mathbf{k},+}^{R}G_{\mathbf{k}^{\prime}+}^{R}G_{\mathbf{k}^{\prime}+}^{A}\mathrm{Re}[G_{\mathbf{k}^{\prime\prime}+}^{A}].
\end{align*}
For systems with 3-fold, 4-fold or 6-fold rotational symmetry, $\sigma_{xy}^{sk}$
is shown to be zero after summing contributions from symmetry-related
points. In systems with 2-fold rotational symmetry, the skew scattering
contribution $\sigma_{xy}^{sk}$ is similarly small due to the factor
$\mathrm{Re}\left[G_{\mathbf{k}''+}^{A}\right]\simeq\frac{\mu-\epsilon_{\mathbf{k}''}}{\eta^{2}+(\epsilon_{\mathbf{k}''}-\mu)^{2}}$.

In conclusion, the extrinsic contributions to the Hall conductivity
in quantum anomalous Hall metals are negligible for systems with rotational
symmetry and time reversal symmetry.

\section{Discussion}\label{sec12}

To conclude, we proposed the $\frac{1}{2}\mathbb{Z}$ topological
invariant to characterize the metallic nature and half quantized Hall
conductivity of topological metals out of an ordinary metallic ferromagnet.
Local symmetry near the Fermi surface safeguards the quantization
of the Hall conductivity. The local symmetry may emerge in the magnetic
structures of topological materials by means of the locality of the
surface states. We have identified two distinct categories of local
unitary and anti-unitary symmetries in proximity to the Fermi surface
of electron states. Numerical calculation provides substantial evidence
to support the existence of the plateau of the half-quantized Hall
conductivity in semi-magnetic heterostructure of topological insulator
$\mathrm{Bi}_{2}\mathrm{Te}_{3}$ and $\mathrm{Bi}_{2}\mathrm{Se}_{3}$
film for $\nu=1$ and topological crystalline insulator SnTe films
for $\nu=2$ and 4.

%\section{Methods}\label{sec11}

\backmatter

\bmhead{Data availability statement}

All data generated or analysed during this study are included in this published article (and its Supplementary Information files).

\bmhead{Acknowledgments}

This work was supported by the Research Grants Council, University
Grants Committee, Hong Kong under Grants No. C7012-21G and No. 17301823
and the National Key R\&D Program of China under Grant No. 2019YFA0308603, and Guangdong Basic and Applied Basic Research Foundation No. 2024A1515010430 and 2023A1515140008.

\bmhead{Author contributions}

S.-Q. S conceived the project. B. F.  performed the theoretical analysis and simulation. B. F. and S.-Q. S. wrote the manuscript with inputs from all authors. All authors contributed to the discussion of the results. 

\bmhead{Competing interests}

The authors declare no competing interests.

\begin{appendices}




\end{appendices}


%\begin{thebibliography}{}
 %  \bibitem{Niemi1983prl} Niemi, A. J. \& Semenoff, G. W. Axial-Anomaly-Induced Fermion Fractionization and Effective Gauge-Theory Actions in Odd-Dimensional Space-Times, \textit{Phys. Rev. Lett.} $\mathbf{51}$,
 %   2077 (1983).

%\end{thebibliography}

%%=============================================%%
%% For presentation purpose, we have included  %%
%% \bigskip command. please ignore this.       %%
%%=============================================%%

%%=============================================%%
%% For presentation purpose, we have included  %%
%% \bigskip command. please ignore this.       %%
%%=============================================%%

\begin{figure}[H]
\includegraphics[width=8cm]{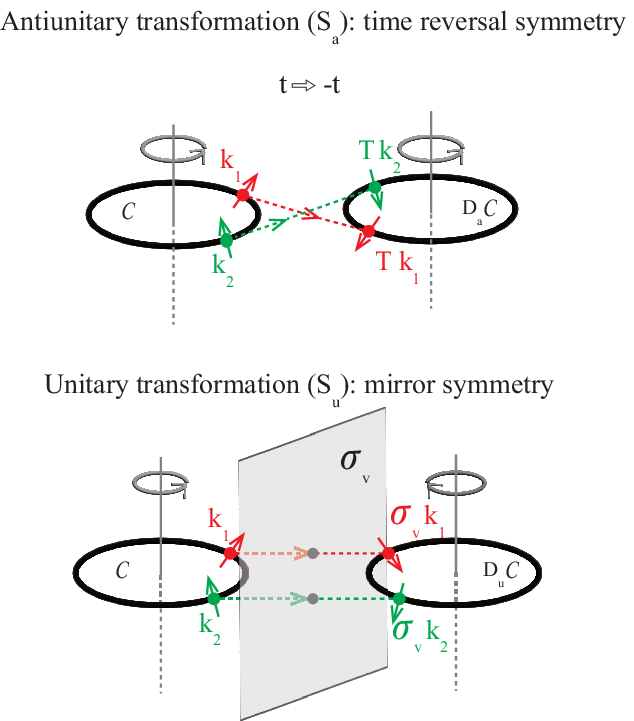}\label{schematic}\caption{\textbf{Schematic diagram illustrating the effects of unitary and anti-unitary
symmetries on a loop integral $C$ of Berry connection $\ointctrclockwiseop_{C}d\mathbf{l}\cdot\mathbf{A}(\mathbf{k}).$}
The upper panel depicts time-reversal symmetry ($T$), and the bottom
panel shows mirror symmetry ($\sigma_{v}$). Under a mirror transformation,
the direction of the loop integral is reversed, whereas, under a time-reversal
transformation, the direction is preserved. The arrows on the states
denote the corresponding spin orientation. As a pseudo-vector or axial
vector, the spin's behavior varies with the symmetry: under mirror
symmetry, its component parallel to the mirror plane is inverted,
while the component perpendicular to the plane remains unchanged.
Under time-reversal symmetry, the entire spin orientation is flipped.}
\end{figure}

\begin{figure}[H]
\includegraphics[width=0.9\textwidth]{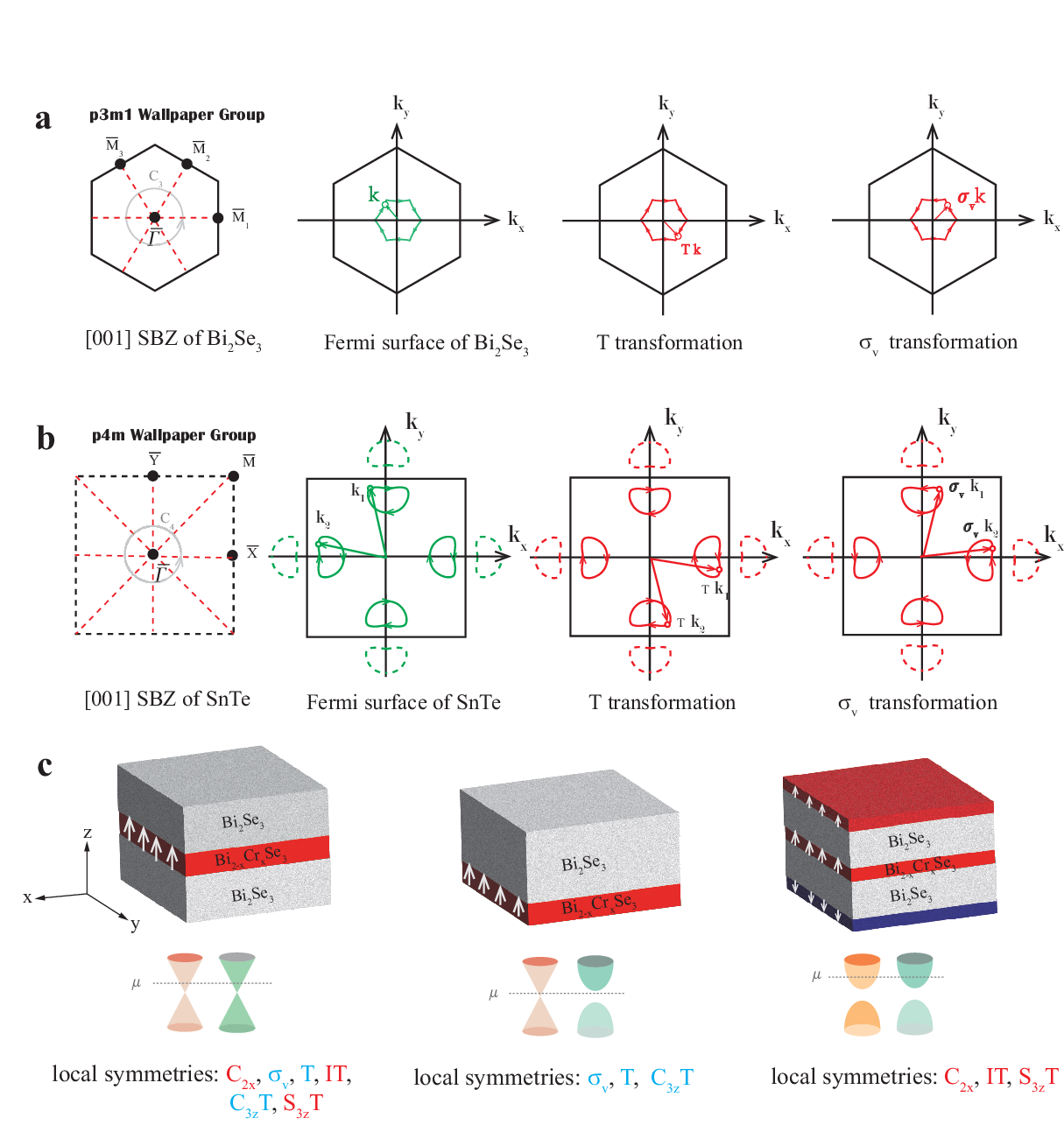}

\caption{\textbf{Local symmetries in the surface states of three-dimensional topological insulator films.} Surface Brillouin zone and transformation after anti-unitary (time
reversal transformation $T$) and unitary symmetry operations (vertical
mirror transformation $\sigma_{v}$ along the y-direction) of the
Fermi surface loop integration for (a) $Bi_{2}Se_{3}$ and (b) $SnTe$
. The black lines indicate the boundary of the Brillouin zone, the
solid dots indicate time-reversal-invariant points, and the (red)
dashed lines indicate mirror lines. (c) $\mathrm{Bi}_{2}\mathrm{Se}_{3}$
film with different magnetic doping scenarios to illustrate the local
symmetries at the Fermi surface $\mu$. Orange cones denote the top
surface states while green cones represent the bottom surface states.
(Left panel) Magnetic doping in the middle layers. (Middel panel)
Semimagnetic film with magnetic doping on bottom surface. (Right panel)
Axion state with magnetic doping applied to both surfaces, where the
exchange field points to opposite directions.The local symmetries
represented in blue map the surface states onto the same surface,
while those marked in red indicate symmetries that map one surface
to another.}
\end{figure}

\begin{figure}[H]
    \centering
    \includegraphics[width=0.9\textwidth]{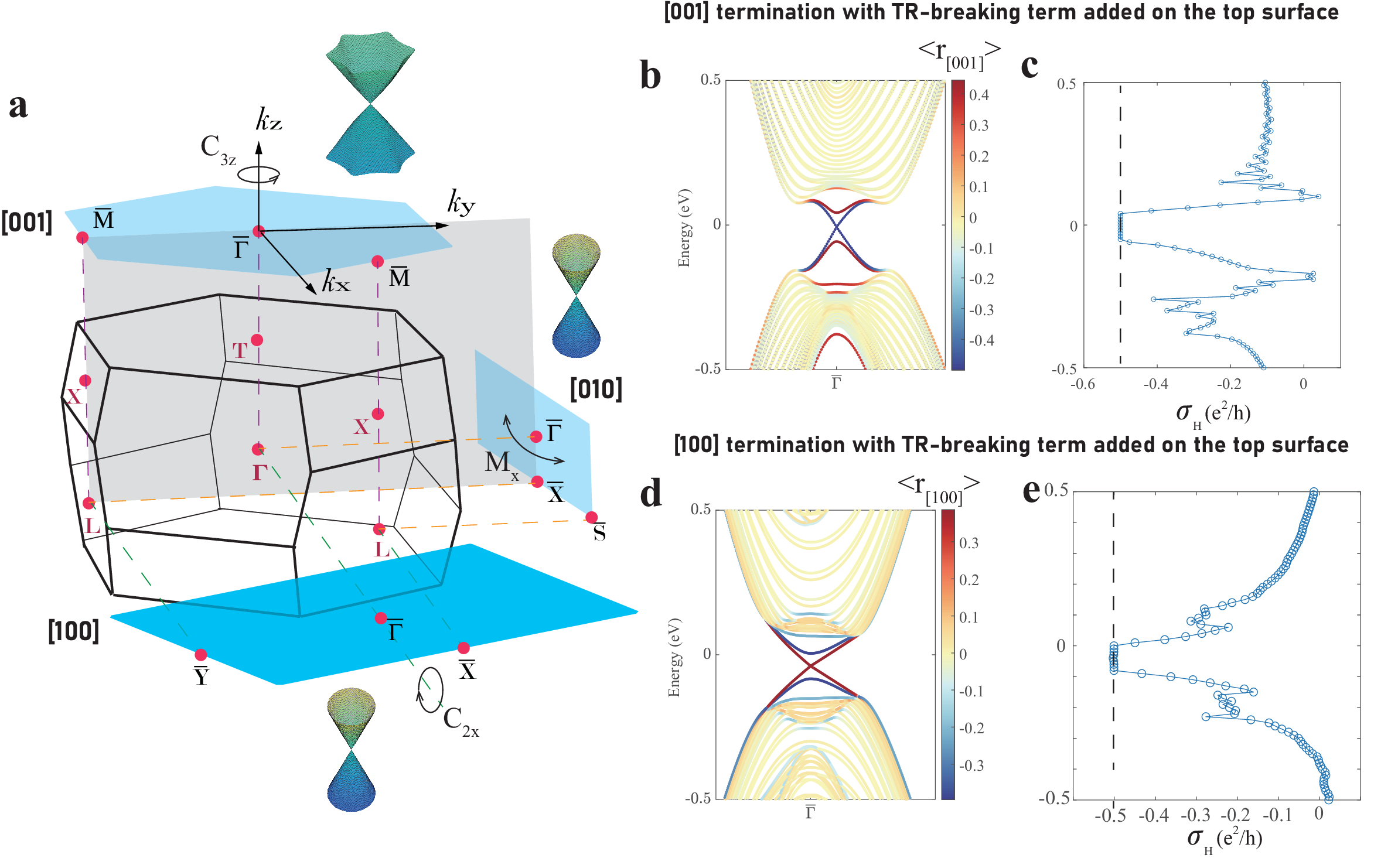}

    \caption{\textbf{The half-quantized Hall phase in strong topological insulators.} (a) Illustration of high-symmetry points within the 3D Brillouin zone
and the corresponding projected surface Brillouin zones for the [001],
[100] and [010] surfaces of the rhombohedral structure, including
the locations of Dirac surface states. (b) Slab band structure around
the $\bar{\Gamma}$ and (c) the Hall conductivity $\sigma_{H}$ as
a function of chemical potential $\mu$ for [001] termination. (f)
The slab band structure around the $\bar{\Gamma}$ and (g) the Hall
conductivity $\sigma_{H}$ as a function of chemical potential $\mu$
for [100] termination. Parameters: $\frac{\mu_{B}}{2}g_{1z}B_{z}=0.05\mathrm{eV}$for
panels (b) and (c); $\frac{\mu_{B}}{2}g_{2x}B_{x}=0.05\mathrm{eV}$
for panels (f) and (g). The tight-binding model and its parameters
are the same as those described in Ref. \citep{fu2024half}.
}
\end{figure}

\begin{figure}[H]
    \centering
    \includegraphics[width=0.9\textwidth]{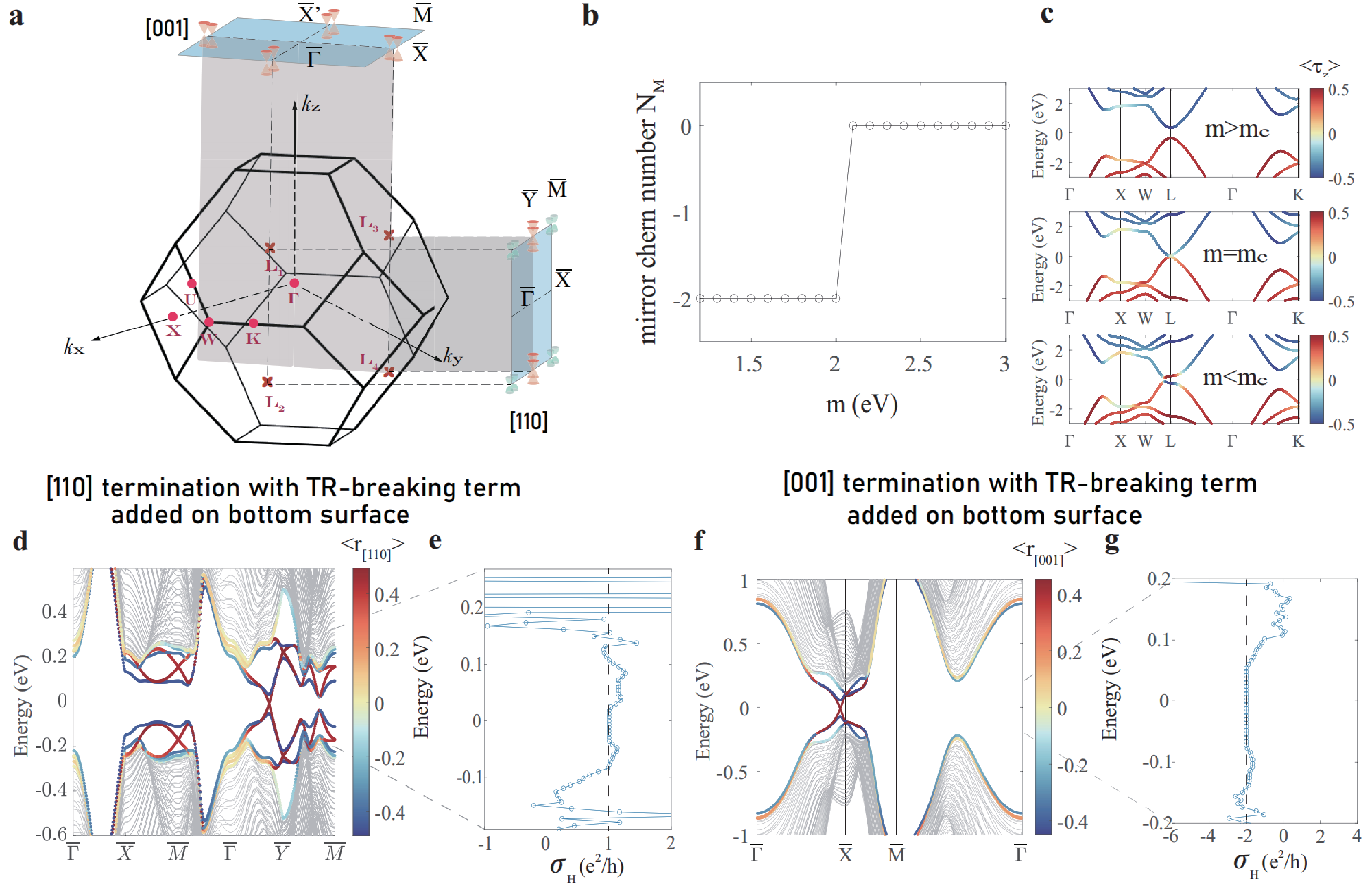}

    \caption{ \textbf{The half-quantized Hall phase in topological crystalline insulator SnTe.}(a) Illustration of high-symmetry points within the 3D Brillouin zone
and the corresponding projected surface Brillouin zones for the [110]
and [001] surfaces of the rock-salt crystal structure, including the
locations of Dirac surface states. (b)The mirror Chern number $N_{M}$
linked to the $\Gamma L_{3}L_{4}$ plane plotted against the parameter
$m$, displaying a step change at $m=m_{c}$. (c) The 3D band structure
along high-symmetry paths for $m>m_{c}$, $m=m_{c}$, and $m<m_{c}$
with color transition from blue to red representing the population
of the wave function at Sn or Te atoms. (d) Slab band structure along
the high-symmetry direction and (e) the Hall conductivity $\sigma_{H}$
as a function of chemical potential $\mu$ for [110] termination.
(f) The slab band structure along high symmetry line and (g) the Hall
conductivity $\sigma_{H}$ as a function of chemical potential $\mu$
for [001] termination. Parameters: $t_{1}=-0.5$, $t_{2}=0.5$, $t_{12}=0.9$
$\lambda_{1}=\lambda_{2}=-0.3$, $m=1.3$ (except for panel (b)),
$\Delta=0.12$ for panels (d), (e), (f), and (g); all values are in
$eV$.
}
\end{figure}

\begin{figure}[H]
    \centering
    \includegraphics[width=0.9\textwidth]{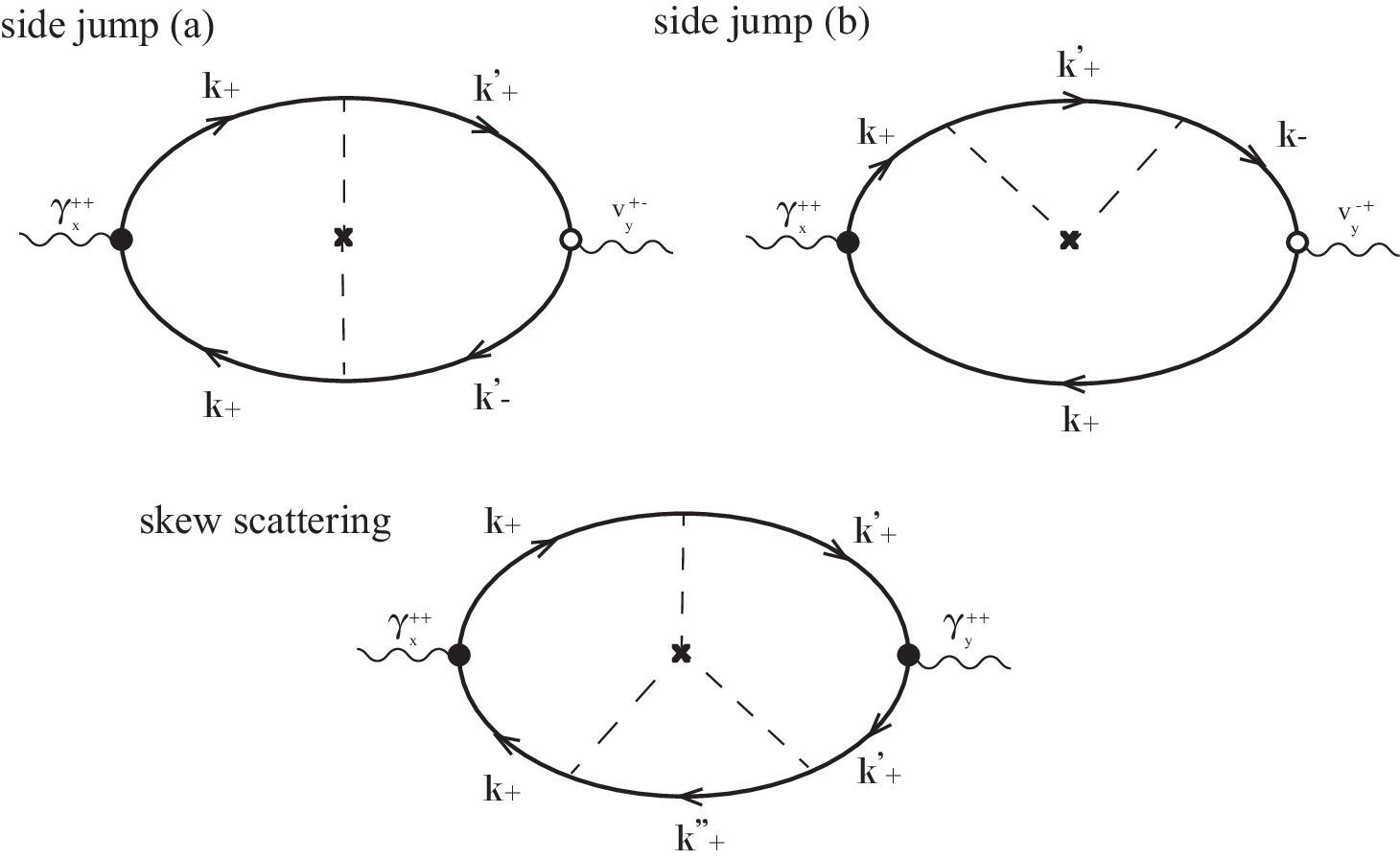}

    \caption{\textbf{Diagrammatic representation of extrinsic contributions to the anomalous Hall effect.} Upper panel: diagrams illustrating side jump contributions.
Lower panel: diagrams depicting skew scattering contributions. The
'+' and '-' symbols denote the upper and lower bands, respectively.
Dashed lines represent impurity interactions. Empty circles indicate
bare current vertices, while filled circles represent renormalized
current vertices.
}
\end{figure}

\begin{landscape}
\begin{table}
\begin{tabular}{c|c|c|c|c|c|c|c|c|c|c|c|c|c}
\hline 
 & \multicolumn{6}{c|}{Unitary $S_{u}$} & \multicolumn{7}{c}{Anti-unitary $S_{a}$}\tabularnewline
\hline 
 & $I$ & $C_{2x}$ & $C_{nz}$ & $\sigma_{v}$ & $\sigma_{h}$ & $S_{nz}$ & $T$ & $IT$ & $C_{2x}T$ & $C_{nz}T$ & $\sigma_{v}T$ & $\sigma_{h}T$ & $S_{nz}T$\tabularnewline
\hline 
\hline 
All $\mathbf{k}$ & Nil & $0$ & Nil & $0$ & Nil & Nil & $0$ & $0$ & Nil & $0$ & Nil & $0$ & $0$\tabularnewline
\hline 
$\mathbf{k}=\mathbf{k}_{F}$ & Nil & $\frac{1}{2}\mathbb{Z}$ & Nil & $\frac{1}{2}\mathbb{Z}$ & Nil & Nil & $\frac{1}{2}\mathbb{Z}$ & $\frac{1}{2}\mathbb{Z}$ & Nil & $\frac{1}{2}\mathbb{Z}$ & Nil & $\frac{1}{2}\mathbb{Z}$ & $\frac{1}{2}\mathbb{Z}$\tabularnewline
\hline 
\end{tabular}

\caption{\textbf{ List of symmetry constraint on the intrinsic Hall conductivity $\sigma_{H}$
 by the generators of
magnetic point groups.} "All $\mathbf{k}$" means that the Hamiltonian
for all $\mathbf{k}$ is invariant under the symmetry while "$\mathbf{k}=\mathbf{k}_{F}$"
means that the Hamiltonian along the Fermi surface is invariant under
the symmetry. "Nil" means that there is no symmetry constraint.
$I$ represent the inversion symmetry, which inverts all spatial coordinates.
The notation $C_{n\alpha}$ denotes the n-fold rotational symmetries
about the $\alpha$-axis. The symbol $\sigma_{v}$ represents vertical
mirror symmetry, characterized by a mirror plane that includes the
z-axis. Conversely, $\sigma_{h}$ denotes horizontal mirror symmetry,
featuring a mirror plane that is perpendicular to the z-axis.The notation
$S_{nz}=C_{nz}\sigma_{h}$ defines an $n$-fold improper rotational
symmetry. This operation consists of first performing a $C_{nz}$
rotation about the z-axis, followed by a reflection in a plane that
is perpendicular to the same axis. Lastly, $T$ represents time reversal
symmetry, which involves reversing the direction of time as well as
other relevant physical quantities, such as magnetic fields and momenta.}
\end{table}
\end{landscape}

\begin{landscape}
\begin{table*}
%{\scriptsize{}%
%\begin{longtable}
\begin{tabular}[c]{|c|c|c|c|c|c|c|c|c|c|c|}
\hline 
 & \multicolumn{5}{c|}{{\scriptsize spin $\frac{1}{2}$}} &  & \multicolumn{4}{c|}{{\scriptsize spinless}}\tabularnewline
\hline 
{\scriptsize$n$} & {\scriptsize$m_{o}$} & {\scriptsize$f(\mathbf{k})$} & {\scriptsize$g(\mathbf{k})$} & {\scriptsize examples} & {\scriptsize$\sigma_{xy}[\frac{e^{2}}{2h}\mathrm{sgn}(V_{z})]$} & {\scriptsize$m_{e}$} & {\scriptsize$f(\mathbf{k})$} & {\scriptsize$g(\mathbf{k})$} & {\scriptsize examples} & {\scriptsize$\sigma_{xy}[\frac{e^{2}}{2h}\mathrm{sgn}(V_{z})]$}\tabularnewline
\hline 
\hline 
{\scriptsize 2} & {\scriptsize 1} & {\scriptsize$v_{+}k_{+}+v_{-}k_{-}$} & {\scriptsize 0} &  & {\scriptsize$\mathrm{sgn}(|v_{-}|-|v_{+}|)$} & {\scriptsize 0} & {\scriptsize$v+v_{+}k_{+}^{2}+v_{-}k_{-}^{2}$} & {\scriptsize 0} & {\scriptsize$\mathrm{SnTe}$\cite{hsieh2012topological}} & {\scriptsize$2\mathrm{sgn}(|v_{-}|-|v_{+}|)$}\tabularnewline
\hline 
{\scriptsize 3} & {\scriptsize 0} & {\scriptsize$v_{+}k_{+}^{3}+v_{-}k_{-}^{3}$} & {\scriptsize Re($v^{\prime}k_{+}^{3}$)} &  & {\scriptsize$3\mathrm{sgn}(|v_{-}|-|v_{+}|)$} & {\scriptsize 0} & {\scriptsize$v+v_{+}k_{+}^{6}+v_{-}k_{-}^{6}$} & {\scriptsize Re($v^{\prime}k_{+}^{3}$)} &  & {\scriptsize$6\mathrm{sgn}(|v_{-}|-|v_{+}|)$}\tabularnewline
\hline 
{\scriptsize 3} & {\scriptsize 1} & {\scriptsize$vk_{-}$} & {\scriptsize Re($v^{\prime}k_{+}^{3}$)} & {\scriptsize$\mathrm{Bi}_{2}\mathrm{Se}(\mathrm{Te})_{3}$\cite{zhang2009topological}} & {\scriptsize$+1$} & {\scriptsize 1} & {\scriptsize$vk_{+}^{2}$} & {\scriptsize Re($v^{\prime}k_{+}^{3}$)} &  & {\scriptsize$-2$}\tabularnewline
\hline 
{\scriptsize 3} & {\scriptsize 2} & {\scriptsize$vk_{+}$} & {\scriptsize Re($v^{\prime}k_{+}^{3}$)} &  & {\scriptsize$-1$} & {\scriptsize 2} & {\scriptsize$vk_{-}^{2}$} & {\scriptsize Re($v^{\prime}k_{+}^{3}$)} &  & {\scriptsize$+2$}\tabularnewline
\hline 
{\scriptsize 4} & {\scriptsize 1} & {\scriptsize$vk_{-}$} & {\scriptsize 0} & {\scriptsize BHZ model\cite{qi2011topological}} & {\scriptsize$+1$} & {\scriptsize 0} & {\scriptsize$v+v_{+}k_{+}^{4}+v_{-}k_{-}^{4}$} & {\scriptsize 0} &  & {\scriptsize$4\mathrm{sgn}(|v_{-}|-|v_{+}|)$}\tabularnewline
\hline 
{\scriptsize 4} & {\scriptsize 3} & {\scriptsize$vk_{+}$} & {\scriptsize 0} &  & {\scriptsize$-1$} & {\scriptsize 2} & {\scriptsize$v_{+}k_{+}^{2}+v_{-}k_{-}^{2}$} & {\scriptsize 0} & {\scriptsize Fu model\cite{fu2011topological}} & {\scriptsize$2\mathrm{sgn}(|v_{-}|-|v_{+}|)$}\tabularnewline
\hline 
{\scriptsize 6} & {\scriptsize 1} & {\scriptsize$vk_{+}$} & {\scriptsize 0} &  & {\scriptsize$-1$} & {\scriptsize 0} & {\scriptsize$v+v_{+}k_{+}^{6}+v_{-}k_{-}^{6}$} & {\scriptsize 0} &  & {\scriptsize$6\mathrm{sgn}(|v_{-}|-|v_{+}|)$}\tabularnewline
\hline 
{\scriptsize 6} & {\scriptsize 3} & {\scriptsize$v_{+}k_{+}^{3}+v_{-}k_{-}^{3}$} & {\scriptsize 0} &  & {\scriptsize$3\mathrm{sgn}(|v_{-}|-|v_{+}|)$} & {\scriptsize 2} & {\scriptsize$vk_{+}^{2}$} & {\scriptsize 0} &  & {\scriptsize$-2$}\tabularnewline
\hline 
{\scriptsize 6} & {\scriptsize 5} & {\scriptsize$vk_{-}$} & {\scriptsize 0} &  & {\scriptsize$-1$} & {\scriptsize 4} & {\scriptsize$vk_{-}^{2}$} & {\scriptsize 0} &  & {\scriptsize$+2$}\tabularnewline
\hline 
\end{tabular}
%\end{longtable}}{\scriptsize\par}

\caption{\textbf{Summary of the classification of the gapless surface states under $C_{n}$ symmetry and time reversal symmetry $T$ in topological insulator films.} The
first column specifies the order of rotation, denoted by $n$. The
second column represent $m_{o}=\mathrm{mod}(2l+1,n)$ with $l$ as
the orbital angular momentum for the first basis state. Columns three
through five represent the symmetry allowed functions $f(\mathbf{k})$
and $g(\mathbf{k})$ at the lowest order of $\mathbf{k}$ (if the
lowest order is $\mathbf{k}$-independent, we truncate to the second
lowest order), and the examples for spin $1/2$ system. Column six
indicates half quantized Hall conductivity corresponding to each case.
The subsequent five columns provide equivalent information for spinless
system. The subsequent five columns provide equivalent information
for spinless system with $m_{e}=\mathrm{mod}(2l,n)$. The coefficients
$v,v_{\pm}$ within $f(\mathbf{k})$ are complex. }
\end{table*}
\end{landscape}

\bibliographystyle{apsrev4-1}
\bibliography{refer}

\end{document}